\def\reals{{\dl R}}
\newcommand{\aim}    {\mbox{$\alpha^{(i)}_m$}}
\newcommand{\asi}    {apparent singularity}
\newcommand{\asis}   {apparent singularities}
\newcommand{\be}     {\begin{equation}}
\newcommand{\bfe}    {{\bf1}}
\newcommand{\cdim}   {conformal dimension}
\newcommand{\cft}    {conformal field theory}
\newcommand{\cfts}   {conformal field theories}
\newcommand{\cijk}   {\mbox{$C_{ij}^{\;\;\;k}$}}
\newcommand{\cgc}    {Clebsch\hy Gor\-dan coefficient}
\newcommand{\chil}   {\mbox{$\chi^{}_{\Lambda}$}}
\newcommand{\chilb}  {\mbox{$\bar\chi^{}_{\Lambda}$}}
\newcommand{\complex}{{\dl C}}
\newcommand{\deq}    {differential equation}
\newcommand{\ee}     {\end{equation}}
\newcommand{\emt}    {ener\-gy\hy mo\-men\-tum tensor}
\newcommand{\fabc}   {\mbox{$f^{ab}_{\:\:\:\:c}$}}
\newcommand{\findim} {fi\-ni\-te-di\-men\-si\-o\-nal}
\newcommand{\four}   {four-point function}
\newcommand{\frc}    {fusion rule coefficient}
\newcommand{\furu}   {fusion rule}
\newcommand{\g}      {\begin{picture}(5.6,4.2)\put(-0.2,0.7){\sf o}
                     \put(1,-2.35){\sf J} \end{picture}}
\newcommand{\ghat}   {\mbox{$\hat{\g}$}}
\newcommand{\hw}     {highest weight}
\newcommand{\hwm}    {highest weight module}
\newcommand{\hy}     {$\mbox{-\hspace{-.66 mm}-}$}
\newcommand{\id}     {{\sl id}}
\newcommand{\ihwm}   {irredu\-cib\-le highest weight module}
\newcommand{\ii}     {{\rm i}}
\newcommand{\infdim} {in\-fi\-ni\-te-di\-men\-si\-o\-nal}
\newcommand{\jf}     {J.\ Fuchs}
\newcommand{\jkil}   {[\!\!{\scriptstyle\begin{array}{c}{\scriptstyle jk}
                     \\[-1.7 mm]{\scriptstyle il} \end{array}}\!\!]}
\def\Jo       {J_\circ}

\def\kze      {Knizh\-nik\hy Za\-mo\-lod\-chi\-kov equation}

\newcommand{\lrc}    {Litt\-le\-wood\hy Ri\-chard\-son coefficient}
\newcommand{\matrx}[2] {\left( \begin{array}{#1} #2 \end{array}\right)}
\newcommand{\nijk}   {\mbox{${\cal N}_{ij}^{\ \;k}$}}
\newcommand{\nikm}   {\mbox{${\cal N}_{ik}^{\;\;\;m}$}}
\newcommand{\njik}   {\mbox{${\cal N}_{ji}^{\;\;\;k}$}}

\newcommand{\njlk}   {\mbox{${\cal N}_{jl}^{\;\;\;k}$}}
\newcommand{\nklm}   {\mbox{${\cal N}_{kl}^{\;\;\;m}$}}
\newcommand{\nlll}{\mbox{${\cal N}_{\!\Lambda\Lambda'}^{\;\;\;\Lambda''}$}}
\newcommand{\nlllb}{\mbox{$\overline{\cal N}{}_{\!\!\Lambda\Lambda'}
                     ^{\;\;\;\Lambda''}$}}
\newcommand{\one}    {\mbox{\small $1\!\!$}1}
\newcommand{\onedim} {one-di\-men\-si\-o\-nal}
\newcommand{\onehalf}{\mbox{$\frac{1}{2}$}}
\newcommand{\opa}    {operator product algebra}
\newcommand{\opc}    {operator product coefficient}
\newcommand{\poleq}  {polynomial equation}
\newcommand{\qdim}   {quantum dimension}
\newcommand{\qft}    {quantum field theory}
\newcommand{\qg}     {quantum group}
\newcommand{\rcft}   {rational conformal field theory}
\newcommand{\rcfts}  {rational conformal field theories}

\newcommand{\subsect}[1]{\section{#1} \setcounter{equation}{0} }
\newcommand{\suse}   {superselection sector}
\newcommand{\threepf}{three-point function}

\newcommand{\tta}    {\mbox{$\theta$}}
\newcommand{\twodim} {two-di\-men\-si\-o\-nal}
\newcommand{\wzw}    {WZW}
\newcommand{\WZW}    {Wess\hy Zu\-mino\hy Wit\-ten}
\newcommand{\wzwm}   {WZW model}
\newcommand{\wzwt}   {WZW theory}
\newcommand{\wzwts}  {WZW theories}
\newcommand{\ybe}    {Yang\hy Bax\-ter equation}
\newcommand{\zet} {{\dl Z}}

\newcommand{\zetplus}{\mbox{$\zet_{>0}$}}
\newcommand{\zetpluso} {\mbox{$\zet_{\geq 0}$}}
\newcommand{\zetplusO} {\zet_{\geq 0}}
\font\tendl=msym10  scaled \magstep1                 
\font\tengl=eufm10  scaled \magstep1                 
\font\sevendl=msym7 scaled \magstep1
\font\fivedl=msym5 scaled \magstep1
\font\sevengl=eufm7 scaled \magstep1
\font\fivegl=eufm5 scaled \magstep1
\newfam\dlfam \def\dl{\fam\dlfam\tendl}          
\textfont\dlfam=\tendl \scriptfont\dlfam=\sevendl
\scriptscriptfont\dlfam=\fivedl
\newfam\glfam           
\textfont\glfam=\tengl \scriptfont\glfam=\sevengl
\scriptscriptfont\glfam=\fivegl
\def\ifundefined#1{\expandafter\ifx\csname#1\endcsname\relax}
\makeatletter \ifundefined{new@mathgroup} {} \else
    \new@mathgroup\sffam
    \new@internalmathalphabet\mathsf\sffam{cmss}{m}{n}
    \def\psf{\fontfamily\sfdefault \fontseries\default@series
        \fontshape\default@shape\selectfont\mathsf}
\fi

\def\citen#1{\if@filesw \immediate\write \@auxout {\string\citation{#1}}\fi%
\@tempcntb\m@ne \let\@h@ld\relax \def\@citea{}%
\@for \@citeb:=#1\do {\@ifundefined {b@\@citeb}%
    {\@h@ld\@citea\@tempcntb\m@ne{\bf ?}%
    \@warning {Citation `\@citeb ' on page \thepage \space undefined}}%
    {\@tempcnta\@tempcntb \advance\@tempcnta\@ne
    \setbox\z@\hbox\bgroup\ifcat0\csname b@\@citeb \endcsname \relax
       \egroup \@tempcntb\number\csname b@\@citeb \endcsname \relax
       \else \egroup \@tempcntb\m@ne \fi \ifnum\@tempcnta=\@tempcntb
       \ifx\@h@ld\relax \edef \@h@ld{\@citea\csname b@\@citeb\endcsname}%
       \else \edef\@h@ld{\hbox{--}\penalty\@highpenalty
	      \csname b@\@citeb\endcsname}\fi
    \else \@h@ld\@citea\csname b@\@citeb \endcsname \let\@h@ld\relax \fi}%
 \def\@citea{,\penalty\@highpenalty\hskip.13em plus.13em minus.13em}}\@h@ld}
\def\@citex[#1]#2{\@cite{\citen{#2}}{#1}}%
\def\@cite#1#2{\leavevmode\unskip
  \ifnum\lastpenalty=\z@\penalty\@highpenalty\fi%
  \ [{\multiply\@highpenalty 3 #1
  \if@tempswa,\penalty\@highpenalty\ #2\fi}]}   %
\makeatother 

   \newcommand{\wb}{\,\linebreak[0]}
   \newcommand{\wB}       {$\,$\wb}
   \newcommand{\J}[1]     {{{#1}}\vyp}
   \newcommand{\JJ}[1]    {{{#1}}\vyp}
   \newcommand{\Bi}[1]    {\bibitem{#1}}
   \newcommand{\bi}[1]    {\bibitem{#1}}
   \newcommand{\Prep}[2]  {preprint {#1}}
   \newcommand{\PRep}[2]  {preprint {#1}}
   
   \newcommand{\Erra}[3]  {\,[{\em ibid.}\ {#1} ({#2}) {#3}, {\em Erratum}]}
   \newcommand{\BOOK}[4]  {{\em #1\/} ({#2}, {#3} {#4})}
   \newcommand{\inBO}[7]  {in:\ {\em #1}, {#2}\ ({#3}, {#4} {#5}), p.\ {#6}}
   
   \newcommand{\vyp}[4]   {\ {#1} ({#2}) {#3}}
   \newcommand{\vypf}[5]  {\ {#1} [FS{#2}] ({#3}) {#4}}
   \newcommand{\adma} {Adv.\wb Math.}

   \newcommand{\anop} {Ann.\wb Phys.}

   \newcommand{\comp} {Com\-mun.\wb Math.\wb Phys.}

   \newcommand{\ctpa} {Com\-mun.\wb Theor.\wb Phys.\ (Allahabad)}
   \newcommand{\duki}{Duke\wB Math.\wb J.\ (Int.\wb Mat.\wb Res.\wb Notes)}
   \newcommand{\ijmb} {Int.\wb J.\wb Mod.\wb Phys.\ B}
   \newcommand{\ijmp} {Int.\wb J.\wb Mod.\wb Phys.\ A}
   \newcommand{\inma} {Invent.\wb math.}
   \newcommand{\jetl} {Sov.\wb Phys.\wB JETP\wB Lett.}
   \newcommand{\jetp} {Sov.\wb Phys.\wB JETP}

   \newcommand{\jomp} {J.\wb Math.\wb Phys.}
   \newcommand{\jopa} {J.\wb Phys.\ A}
   \newcommand{\josm} {J.\wb Sov.\wb Math.}
   \newcommand{\jram} {J.\wB rei\-ne\wB an\-gew.\wb Math.}
   \newcommand{\kniz}[2] {\inBO{The Physics and Mathematics of Strings,
              Memorial Volume for V.G.\ Knizhnik} {L. Brink et al., eds.}
              \WS\Si{1990} {{#1}}{{#2}} }
   \newcommand{\lemp} {Lett.\wb Math.\wb Phys.}

   \newcommand{\mqft}[2] {\inBO{{\rm Proceedings of the} International
              Colloquium on Modern Quantum Field Theory {\rm (Bombay,
              January 1990)}} {} \WS\Si{1991} {{#1}}{{#2}} }
   \newcommand{\mpla} {Mod.\wb Phys.\wb Lett.\ A}
   
   \newcommand{\npbf} {Nucl.\wb Phys.\ B\vypf}
   \newcommand{\npbp} {Nucl.\wb Phys.\ B (Proc.\wb Suppl.)}
   \newcommand{\nspq}[2] {\inBO{{\rm 1991 Carg\`ese Lectures on} New
              Symmetry Principles in Quantum Field Theory} {J.\
              Fr\"ohlich et al., eds.}{\PL}{New York}{1992} {{#1}}{{#2}}}
   \newcommand{\nuci} {Nuovo\wB Cim.}
   \newcommand{\nupb} {Nucl.\wb Phys.\ B}
   
   \newcommand{\phlb} {Phys.\wb Lett.\ B}

   \newcommand{\phrd} {Phys.\wb Rev.\ D}
   \newcommand{\phre} {Phys.\wb Rev.}
   \newcommand{\phrl} {Phys.\wb Rev.\wb Lett.}
   \newcommand{\prep} {Phys.\wb Rep.}
   
   \newcommand{\rims} {Publ.\wB RIMS}
   \newcommand{\rmap} {Rev.\wb Math.\wb Phys.}

   \newcommand{\slnm} {Sprin\-ger Lecture Notes in Mathematics}
   \newcommand{\slnp} {Sprin\-ger Lecture Notes in Physics}

   \newcommand{\Suse} [2] {\inBO{The Algebraic Theory of Superselection
              Sectors.\ Introduction and Recent Results} {D. Kastler,
              ed.} \WS\Si{1990} {{#1}}{{#2}} }

   \newcommand{\AP}     {{Academic Press}}

   \newcommand{\BIR}    {{Birk\-h\"au\-ser}}
   \newcommand{\CUP}    {{Cambridge University Press}}

   \newcommand{\JW}     {{John Wiley}}

   \newcommand{\PL}     {{Plenum}}
   
   \newcommand{\SV}     {{Sprin\-ger Verlag}}
   
   \newcommand{\WS}     {{World Scientific}}
   \newcommand{\Be}     {{Berlin}}
   \newcommand{\Ca}     {{Cambridge}}
   \newcommand{\NY}     {{New York}}

   \newcommand{\Si}     {{Singapore}}
\long\def\query#1{%
\hskip 0pt{\vadjust{\everypar={}\small\vtop to 0pt{\hbox{}%
\vskip -13pt\rlap{\hbox to 47.5pc{\hfil{\vtop{\hsize=8pc\tolerance=6000%
\hfuzz=.5pc\rightskip=0pt plus 3em\noindent#1}}}}\vss}}}}%
\def\dimn   {dimension}
\def\lagi   {Lan\-dau\hy Ginz\-burg}
\def\qg        {quantum group}
\def\Rep    {Representation}
\newcommand{\labl}[1]{\label{#1}\ee}
\newcommand{\mpcir}[6] {\multiput(#1,#2)(#3,#4){#5}{\circle{#6}}}
\newcommand{\mpcis}[6] {\multiput(#1,#2)(#3,#4){#5}{\circle*{#6}}}
\newcommand{\mplin}[8] {\multiput(#1,#2)(#3,#4){#5}{\line(#6,#7){#8}}}
\newcommand{\mpliv}[9] {\multiput(#1,#2)(#3,#4){#5}{\line(#6,#7){#8}}
                       \multiput(#1,#2)(#3,#4){#5}{\vector(#6,#7){#9}}}
\newcommand{\pucir}[3] {\put(#1,#2){\circle{#3}}}

\newcommand{\pulin}[5] {\put(#1,#2){\line(#3,#4){#5}}}
\newcommand{\puliv}[6] {\put(#1,#2){\line(#3,#4){#5}}
                       \put(#1,#2){\vector(#3,#4){#6}}}
\newcommand{\putfs}[3] {\put(#1,#2){{\footnotesize #3}}}
\newcommand{\putss}[3] {\put(#1,#2){{\scriptsize #3}}}
\newcommand{\puvec}[5] {\put(#1,#2){\vector(#3,#4){#5}}}
\newcommand{\piccalenine} {\begin{picture}(160,100)(-6,41)
 \mpcir00{80}035 \mpcir{40}{40}{40}035 \mpcir{60}{20}{40}025 \pucir{80}{80}5
 \mpliv{157.5}0{-80}02{-1}0{75}{45}
 \mpliv{42.5}{40}{40}0210{35}{21} \puliv{62.5}{20}10{35}{21}
 \mpliv{1.8}{1.8}{40}{40}211{36.4}{23}
 \mpliv{78.2}{1.8}{-20}{20}2{-1}1{16.4}{10}
 \puliv{98.2}{21.8}{-1}1{16.4}{10}
 \mpliv{81.8}{78.2}{40}{-40}21{-1}{36.4}{23}
 \mpliv{118.2}{38.2}{-20}{-20}2{-1}{-1}{16.4}{10}
 \puliv{78.2}{38.2}{-1}{-1}{16.4}{10} \end{picture}}
\newcommand{\picslzzfg} {\begin{picture}(160,86)(0,-11)
 \putfs{-2}{-11}{0\hskip 15mm$\la_{(1)}$\hskip 12mm$2\la_{(1)}$}
 \putfs1{24}{$\la_{(2)}$\hskip 22.3mm$\la_{(1)}+\la_{(2)}$}
 \putfs{43}{58}{$2\la_{(2)}$}
 \mpcir00{50}035 \mpcir{25}{25}{50}025   \pucir{50}{50}5
 \mpliv{2.5}0{50}0210{45}{27} \puliv{27.5}{25}10{45}{27}
 \mpliv{23.2}{23.2}{25}{25}2{-1}{-1}{21.4}{13}
 \puliv{73.2}{23.2}{-1}{-1}{21.4}{13}
 \mpliv{98.2}{1.8}{-25}{25}2{-1}1{21.4}{13}
 \puliv{48.2}{1.8}{-1}1{21.4}{13} \end{picture}}
\def\picsldzfg {\begin{picture}(160,125)(0,-11)
 \mpcir00{50}035 \mpcir0{50}{100}025 \mpcir0{100}{50}035
 \mpcir{32}{70}{36}{-40}25
 \mpliv{48.5}{2.3}{19}{29.8}2{-1}4{16.3}{12}
 \mpliv{29.7}{69.7}{68}{-20.8}2{-3}{-2}{27.8}{15}
 \mpliv{48.8}{97.4}{19.2}{-70}2{-2}{-3}{16.9}{10}
 \mpliv{1.7}{48.2}{62.2}{12}24{-1}{34.2}{22}
 \mplin{33.8}{68.2}{8.7}{-31}24{-1}{22.5}
 \putss{-2}{108}0 \putss{44}{108}{$\la_{(3)}$}
 \putss{93}{108}{$2\la_{(3)}$} \putss{-6}{-11}{$2\la_{(1)}$}
 \putss{32}{-11}{$\la_{(1)}\!+\!\la_{(2)}$} \putss{93}{-11}{$2\la_{(2)}$}
 \putss{-22}{48.5}{$\la_{(1)}$} \putss{105}{48.5}{$\la_{(2)}\!+\!\la_{(3)}$}
 \putss{6}{83.8}{$\la_{(1)}$} \putss{8.6}{73.7}{$+\!\la_{(3)}$}
 \putss{70}{23}{$\la_{(2)}$}
 \mpliv{2.5}0{50}0210{45}{27} \mpliv{97.5}{100}{-50}02{-1}0{45}{27}
 \mpliv0{97.5}0{-50}20{-1}{45}{27} \mpliv{100}{2.5}0{50}201{45}{27}
\end{picture}}
\newcommand{\picslzkfg} {\begin{picture}(160,133)(30,-11)
 \putss{-2.3}{-13}0 \putss{24}{-13}{$\la_{(1)}$}
 \putss{202.2}{-13}{$k\la_{(1)}$} \putss{-5}{16}{$\la_{(2)}$}
 \putss{95.7}{115}{$k\la_{(2)}$}
 \mpcir00{30}045 \mpcir{15}{15}{30}035   \mpcir{30}{30}{30}025
 \pucir{45}{45}5 \mpcir{180}0{-15}{15}75 \mpcir{210}0{-15}{15}85
 \mpcis{125}05061\mpcis{61}{61}3361
 \mpliv{2.5}0{30}0410{25}{16}      \mpliv{17.5}{15}{30}0310{25}{16}
 \mpliv{32.5}{30}{30}0210{25}{16}  \puliv{47.5}{45}10{25}{16}
 \mpliv{157.5}0{-15}{15}610{20}{13}\mpliv{182.5}0{-15}{15}710{25}{16}
 \mpliv{13.2}{13.2}{15}{15}4{-1}{-1}{11.4}{7.3}
 \mpliv{43.2}{13.2}{15}{15}3{-1}{-1}{11.4}{7.3}
 \mpliv{73.2}{13.2}{15}{15}2{-1}{-1}{11.4}{7.3}
 \puliv{103.2}{13.2}{-1}{-1}{11.4}{7.3}
 \mpliv{178.2}{1.8}{-15}{15}6{-1}1{11.4}{7.3}
 \mpliv{208.2}{1.8}{-15}{15}7{-1}1{11.4}{7.3}
 \puliv{28.2}{1.8}{-1}1{11.4}{7.3}
 \mpliv{58.2}{1.8}{-15}{15}2{-1}1{11.4}{7.3}
 \mpliv{88.2}{1.8}{-15}{15}3{-1}1{11.4}{7.3}
 \mpliv{193.2}{13.2}{-15}{15}7{-1}{-1}{11.4}{7.3}
 \mpliv{163.2}{12.5}{-15}{15}6{-1}{-1}{10}{6}
\end{picture}}
 \newcommand{\picesixe}
{\begin{picture}(160,80)(-8,-25)
\mpcir{2.5}{10}{30}055 \mpcir{62.5}{40}0{30}25
\mplin5{10}{30}0410{25} \mplin{62.5}{12.5}0{30}201{25}
\putss{.8}{-4}1 \putss{30.8}{-4}2 \putss{60.8}{-4}3
\putss{90.8}{-4}2 \putss{120.8}{-4}1 \putss{72}{37.4}2
\putss{72}{67.5}1 \end{picture}}
 \newcommand{\picarbar}
{\begin{picture}(240,18)(-8,7)
\put(209,12.5){\oval(25,10)[t]} \put(209,7.5){\oval(25,10)[b]}
\mpcir{2.5}{10}{30}035 \mpcir{166.5}{10}{30}025 \mpcis{97}{10}7061
\mplin5{10}{30}0310{25} \mplin{139}{10}{30}0210{25}
\pulin{221.5}{7.5}015 \end{picture}}
 \newcommand{\picaerbar}
{\begin{picture}(265,30)(-1,7)
\mplin{35}{10}{30}0310{25} \mpcir{32.5}{10}{30}035
\mpcis{127}{10}7061 \mplin{169}{10}{30}0210{25}
\mpcir{196.5}{10}{30}025 \pulin{251.5}{7.5}015 \pulin{7.5}{7.5}015
\put(239,12.5){\oval(25,10)[t]}\put(239,7.5){\oval(25,10)[b]}
\put(20,12.5){\oval(25,10)[t]} \put(20,7.5){\oval(25,10)[b]}
\end{picture}}
 \newcommand{\picderbar}
{\begin{picture}(265,38)(-4,7)
\mplin{35}{10}{30}0310{25} \mpcir{32.5}{10}{30}035
\mpcis{127}{10}7061 \mplin{169}{10}{30}0210{25}
\mpcir{196.5}{10}{30}025 \mpcir{7}{-2.9}0{25.8}25
\put(30.2,11.2){\line(-2,1){21}} \put(30.2,8.8){\line(-2,-1){21}}
\put(239,12.5){\oval(25,10)[t]} \put(239,7.5){\oval(25,10)[b]}
\put(251.5,7.5){\line(0,1){5}} \end{picture}}
 
\newcommand{\smallpicnijk}[3]
{\begin{picture}(80,50)(0,7) \put(30,10){\line(1,0){35}}
\put(30,10){\line(-1,-1){24.5}} \put(30,10){\line(-1,1){24.5}}
\put(45,10){\vector(1,0){2.5}}
\put(15,25){\vector(1,-1){2.5}} \put(15,-5){\vector(1,1){2.5}}
\put(-1,32){$#1$} \put(-1,-16){$#2$} \put(67,8){$#3$} \end{picture}}
\newcommand{\smallpicnijuk}[3]
{\begin{picture}(80,50)(0,7) \put(30,10){\line(1,0){35}}
\put(30,10){\line(-1,-1){24.5}} \put(30,10){\line(-1,1){24.5}}
\put(50,10){\vector(-1,0){2.5}}
\put(15,25){\vector(1,-1){2.5}} \put(15,-5){\vector(1,1){2.5}}
\put(-1,32){$#1$} \put(-1,-16){$#2$} \put(67,8){$#3$} \end{picture}}
\newcommand{\smallpicnijo}[3]
{\begin{picture}(80,50)(0,7) \multiput(30,10)(3.5,0){10}{\line(1,0){2}}
\put(30,10){\line(-1,-1){24.5}} \put(30,10){\line(-1,1){24.5}}
\put(15,25){\vector(1,-1){2.5}} \put(15,-5){\vector(1,1){2.5}}
\put(-1,32){$#1$} \put(-1,-16){$#2$} \put(67,8){$#3$} \end{picture}}

\newcommand{\smallpicschannel}[5]
{{\begin{picture}(100,50)(0,8) \put(30,10){\line(1,0){35}}
\put(30,10){\line(-1,-1){24.5}} \put(30,10){\line(-1,1){24.5}}
\put(65,10){\line(1,-1){24.5}} \put(65,10){\line(1,1){24.5}}
\put(49,10){\vector(1,0){2.5}}
\put(15,25){\vector(1,-1){2.5}} \put(15,-5){\vector(1,1){2.5}}
\put(80,25){\vector(-1,-1){2.5}} \put(80,-5){\vector(-1,1){2.5}}
\put(-1,32){$#1$} \put(-1,-18){$#2$} \put(44,2){$#3$}
\put(92,32){$#4$} \put(92,-18){$#5$} \end{picture}}}
\newcommand{\smallpictchannel}[5]
{{\begin{picture}(100,50)(0,8)
\puliv{30}{10}10{35}{21} \pulin{30}{10}{-1}{-1}{24.5}
\pulin{65}{10}1{-1}{24.5}\pulin{65}{10}{-2}1{15}
\puliv9{38}2{-1}{36}{21} \put(15,-5){\vector(1,1){2.5}}
\puliv{86}{38}{-2}{-1}{56}{21} \put(80,-5){\vector(-1,1){2.5}}
\put(0,37){$#1$} \put(0,-18){$#2$} \put(46,2){$#3$}
\put(92,37){$#4$}\put(92,-18){$#5$} \end{picture}}}
\newcommand{\smallpicuchannel}[5]
{{\begin{picture}(65,85)(20,8) \put(45,-5){\line(0,1){35}}
\put(45,30){\line(-1,1){24.5}} \put(45,30){\line(1,1){24.5}}
\put(45,-5){\line(-1,-1){24.5}} \put(45,-5){\line(1,-1){24.5}}
\put(45,13){\vector(0,1){2.5}}
\put(30,45){\vector(1,-1){2.5}} \put(60,45){\vector(-1,-1){2.5}}
\put(30,-20){\vector(1,1){2.5}} \put(60,-20){\vector(-1,1){2.5}}
\put(14,54){$#1$} \put(73,54){$#2$} \put(36,12){$#3$}
\put(14,-33){$#4$}\put(73,-33){$#5$} \end{picture}}}
\newcommand{\alg}    {algebra}
\newcommand{\ca}     {\mbox{$\cal A$}}
\newcommand{\calf}{\mbox{$\cal F$}}
\newcommand{\calfzz}{\mbox{${\cal F}(z,\bar z)$}}
\newcommand{\cat}    {\mbox{$\tilde{\cal A}$}}
\newcommand{\chii}   {\mbox{$\chi^{}_i$}}
\newcommand{\cd}[1]  {\mbox{${\cal D}_{#1}$}}
\newcommand{\cn}[1]  {\mbox{${\cal N}_{#1}$}}
\newcommand{\czi}    {\mbox{$\complex\zet_\II$}}
\newcommand{\dm}     {\mbox{$V$}}
\newcommand{\dM}[1]  {\mbox{$V_{#1}$}}
\newcommand{\dsumi}[1]{{\displaystyle\sum_{#1\in I}}}
\newcommand{\dtx}    {\frac\partial{\partial \tilde x_1}}
\newcommand{\dtX}[1] {\partial V(#1)/\partial \tilde x_1}
\newcommand{\dx}[1]  {\frac\partial{\partial x_{#1}}}
\newcommand{\dX}[2]  {\partial V(#2)/\partial x_{#1}}
\newcommand{\dy}     {\mbox{$Y$}}
\newcommand{\fa}     {\mbox{$\cal A$}}
\newcommand{\fra}    {fusion rule algebra}
\newcommand{\frsa}   {fusion rule subalgebra}
\newcommand{\ga}[1]  {\mbox{$\Gamma^{\!{}^\circ}_{\!#1}$}}
\newcommand{\gA}     {\mbox{$\Gamma^{\!{}^\circ}$}}
\newcommand{\gb}[1]  {\mbox{$\Gamma_{\!#1}$}}
\newcommand{\gB}     {\mbox{$\Gamma$}}

\newcommand{\Ii}     {\mbox{$|I|$}}
\newcommand{\II}     {{|I|}}
\newcommand{\irrep}  {irreducible representation}
\newcommand{\jam}    {\mbox{$J^a_m$}}
\renewcommand{\L}      {{\rm L}}
\newcommand{\la}     {{\Lambda}}
\newcommand{\lap}    {{\Lambda'}}
\newcommand{\lapp}   {{\Lambda''}}
\newcommand{\las}    {{\Lambda^+_{}}}
\newcommand{\lla}    {\mbox{${\rm L}_\Lambda$}}
\newcommand{\lle}    {{\Lambda_{(1)}}}
\newcommand{\lles}   {{\Lambda_{(1)}+\Lambda_{(7)}}}
\newcommand{\lls}    {{\Lambda_{(7)}}}
\newcommand{\mm}     {N}
\newcommand{\more}[1]{{}}
\newcommand{\mv}[2]  {\mbox{$\mu_{#1}^{(#2)}$}}
\newcommand{\mV}[1]  {\mbox{$\tilde\mu_{#1}^{(1)}$}}
\newcommand{\mvx}[3] {(x_{#1}-\mu_{#2}^{(#3)})}
\newcommand{\mvX}[1] {(\tx-\tilde\mu_{#1}^{(1)})}
\newcommand{\nv}[2]  {\mbox{$\nu_{#1}^{(#2)}$}}
\newcommand{\pireg}  {\mbox{$\pi_{\rm reg}^{}$}}
\newcommand{\pix}    {\mbox{$\pi_{\rm reg}^{}(x)$}}
\newcommand{\picij} {\begin{picture}(80,50)(0,7) \put(10,10){\line(1,0)
       {60}} \put(25,10){\vector(1,0){2.5}} \put(55,10){\vector(-1,0)
       {2.5}} \put(10,0){$i$} \put(66,0){$j$} \end{picture}}
\newcommand{\pk}[1]  {\mbox{${\cal P}_k(#1)$}}
\newcommand{\proi}[1]{\prod_{#1\in I}}
\newcommand{\proI}[1]{\prod_{#1=0}^{\II-1}}

\newcommand{\proine}[2]{\prod_{\stackrel{\scriptstyle #1=0}{#1\neq#2}}^{\II-1}}
\newcommand{\proinE}[2]{\prod_{#1=0,#1\neq#2}^{\II-1}}
\newcommand{\prom}[1]{\prod_{#1=0}^\mm}
\newcommand{\promne}[2]{\prod_{\stackrel{\scriptstyle #1=0}{#1\neq#2}}^\mm}
\newcommand{\pslz}   {\mbox{{\em PSL}$(2,\zet)$}}
\newcommand{\rep}    {representation}
\newcommand{\rfa}    {rational fusion rule algebra}
\newcommand{\sln}    {\mbox{{\sl sl}$^{}_n$}}
\newcommand{\slnk}   {\mbox{({\sl sl}$^{}_n$)$^{}_k$}}
\newcommand{\slz}    {\mbox{{\em SL}$(2,\zet)$}}
\newcommand{\sx}     {\mbox{$\cal S$}}
\newcommand{\sumi}[1]{\sum_{#1\in I}}
\newcommand{\sumI}[1]{\sum_{#1=0}^{\II-1}}
\newcommand{\sumik}[1]{\sum_{#1\in I_k}}
\newcommand{\sumine}[2]{\sum_{\stackrel{\scriptstyle #1=0}{#1\neq#2}}^{\II-1}}
\newcommand{\summ}[1]{\sum_{#1=0}^\mm}

\newcommand{\ttau}   {${\sf T}_\tau$}
\newcommand{\tx}     {\mbox{$\tilde x_1$}}
\newcommand{\tX}     {\mbox{$\cal T$}}

\newcommand{\vv}[2]  {\mbox{$(v_{#1})^{}_{#2}$}}
\newcommand{\w}      {\mbox{$\cal W$}}
\newcommand{\wo}     {\mbox{${\cal W}_0$}}
\newcommand{\walg}   {$\cal W$-algebra}
\newcommand{\yy}[2]  {\mbox{$Y^{}_{#1#2}$}}
\newcommand{\yz}[2]  {\mbox{$(Y^{-1})^{}_{#1#2}$}}
\def\ibox    {\item[{\tiny$\Box$}\ ]}
\documentstyle[11pt]{article}
\begin{document}
\rightline{\sf hep-th/9306162}
\rightline{\sf June 1993}
\begin{center}{\mbox{ }\\[8 mm]\Large\bf FUSION RULES\\[1 mm] IN CONFORMAL
     FIELD THEORY} \\[16 mm]
     {\large $\quad$ \sc J\"urgen Fuchs~$^\#$} \\[5 mm] NIKHEF-H \\[1 mm]
     Kruislaan 409\\[1 mm]NL\,-\,1098 SJ~~Amsterdam      \\[1 cm] {~}
\end{center}
\begin{quote}{\bf Abstract.}~~Several aspects of fusion rings and
\fra s, and of their manifestations in \twodim\ (conformal) field theory,
are described: diagonalization
and the connection with modular invariance; the presentation in terms
of quotients of polynomial rings; fusion graphs; various strategies
that allow for a partial classification; and the role of the \furu s
in the conformal bootstrap programme.
\end{quote}
\vskip 3.5 cm
   --------------------------- \\[2 mm] $^\#\;$\ Heisenberg fellow
\newpage

\subsect{Fusion rule algebras}

In this paper I describe various aspects of \furu s, or more precisely,
of fusion rings and \fra s. Most of the contents is not entirely new, but
rather a collection of known results, supplemented by examples.
However, the presentation and emphasis is different from
available expositions of the subject. For instance, I describe in
detail the issue of diagonalisation of a \fra, which is related to, but
more fundamental than the Verlinde formula; to make this distinction
explicit, I introduce the notion of a modular \fra, which must not be
confused with the issue of modular invariance in \cft.
I also clarify further the properties of fusion rings that are needed
to represent them as local rings, which necessitates to distinguish
carefully between the fusion rules as a ring over the integers
and as an algebra over the complex numbers, respectively.

Fusion rule \alg s are certain
associative algebras over the complex numbers which arise in various
areas of physics and mathematics, where they describe the possible
couplings among three objects out of some given class.  As examples let
me mention:
\begin{itemize} \ibox The decomposition of
tensor products of \findim\ \rep s of reductive Lie \alg s, of finite
groups, and of associative (bi-)\,algebras, into \irrep s.
\ibox The composition of \suse s in the C$^*$-\alg ic
approach to relativistic \qft\ \cite{frrs,frrs2,doro4,masc}.
\ibox The multiplication of (equivalence classes of) polynomials
in certain quotients of polynomial rings.
\ibox Truncated tensor products of unitary \rep s of quantum groups
with deformation parameter a root of unity \cite{Algs,pasa,fugp2,fuva2}.
\ibox The coupling of primary fields of \walg s in \twodim\ \cft\
\cite{bepz,verl2,mose}.
\end{itemize}
 (For a few further realizations see section 7 below.)\\
If the axioms of a \fra\ are slightly relaxed, one can also describe:
\begin{itemize}
\ibox The multiplication of (classes of) polynomials in any
quotient of a polynomial ring, e.g.\ the ring of chiral primary
fields in $N=2$ super\cft\ \cite{levw,gepn9}.
\ibox Operator products in topological field theory
\cite{witt22,divv3}.  \end{itemize}
 In the present paper the main interest is in the realization of
\furu s in \cft, but to motivate the concept of fusion rings and
\fra s it seems to me most convenient to start
with the first example in the above list, i.e.\ with the
decomposition of tensor products of \findim\ \rep s, respectively
modules, of a reductive Lie \alg.  Thus let \g\ denote a simple Lie \alg\
(the generalization to arbitrary reductive Lie \alg s will be
immediate).  Any
\findim\ module of \g\ and any tensor product of such modules is fully
reducible, i.e.\ the direct sum of irreducible modules.  The \findim\
irreducible modules are \hwm s labelled by a dominant integral \hw\ $\la$
of \g; I denote them by \lla, and their Kronecker tensor product and its
decomposition into irreducible modules by
  \be \lla\times \L_\lap =\bigoplus_\lapp \nlll\, \L_\lapp. \labl a
(Here and below I use, for $V$ a vector space and $n\in\zetpluso$,
the short-hand notation $\,nV$ in place of
$\;\bigoplus_{a=1}^nV^{(a)}_{}$ with $V^{(a)}\cong V$.) \\[2 mm]
Let me recall a few well-known properties of such tensor products:
\\(a) The addition $\oplus$ and product $\times$ are commutative,
associative, and distributive.
\\(b) By definition, the numbers \nlll\ are non-negative integers.
\\(c) The number of \hwm s $\L_\la$ is infinite; but
for fixed $\la$ and $\lap$, \nlll\ is nonzero only for a finite
number of \hw s $\lapp.$
\\(d) The \hwm\ with \hw\ $\la=0$ (the trivial \onedim\ module) acts
as the identity, i.e. $\lla\times\L_0=\lla$, or in other words
  \be  {\cal N}_{\la\,0}^{\;\ \lap}=\delta^{}_{\la,\lap}\,,  \labl b
for any dominant integral \hw\ $\la$.
\\(e) To any module \lla\ there exists a unique conjugate module $\L_
\la^+$ which is again a \findim\ \hwm\ (namely $\L_\la^+=\L_\las^{}$,
with $\las$ being minus the lowest weight of \lla), such that $(\L_\la^
+)^{+_{}}_{}=\lla$, that $\L_0$ appears in $\lla\star\L_\la^+$ precisely
once, i.e.
  \be  {\cal N}_{\la\lap}^{\ \ \;0}=\delta_{\lap,\las},    \labl c
and that the tensor product of conjugate modules is conjugate to
the tensor product of the modules themselves, in the sense that
  \be  {\cal N}_{\las\,\lap^+}^{\ \ \ \ \lapp^+}=\nlll.     \labl d
(f) The trivial module $\L_0$ is self-conjugate.

It is sometimes convenient to describe
the collection of all \lla\ of \g\ and their direct sums
from a category theoretic point of view. This collection constitutes a
category with the objects being the \findim\ \g-modules and the morphisms
(arrows) being the intertwiners between them, i.e.\ the maps $T$ that
make the diagram
  \be  \begin{array}{rcl}
  \lla & \stackrel{\pi^{}_\la(x)}\longrightarrow & \lla \\[2 mm]
  {\scriptstyle T}\,\downarrow\ && \ \downarrow\,{\scriptstyle T}\\[2 mm]
  \L_\lap & \stackrel{\pi^{}_\lap(x)}\longrightarrow & \L_\lap
  \end{array}\ee
(with $\pi_\la$ the \g-\rep\ corresponding to the module \lla)
commutative for all $x\in\g$. The Kronecker tensor
product endows this category with a tensor product functor which
satisfies obvious associativity and commutativity constraints
and which allows for the notion of an identity object and dual (conjugate)
objects;
the commutativity isomorphisms are involutive, and hence the category
has the structure of a rigid tensor (or monoidal) category.

The intertwiners between $\lla\times\L_\lap$ and its irreducible
submodules describe the Clebsch\hy Gor\-dan decomposition of the product;
that is, for $v^\la_\lambda\in\lla$, etc., the tensor product states obey
  \be  v^\la_\lambda\otimes v^\lap_{\lambda'}=\sum_{\lapp,a}
  {\cal C}^{\la\lap;\lapp,a}_{\lambda\lambda'} v^{\lapp,a}_
  {\lambda+\lambda'} \,. \labl e
In this setting, the tensor product coefficients \nlll\
describe the basis independent contents of the
Clebsch\hy Gordan decomposition, namely the dimensionality of the
intertwiner spaces; thus the degeneracy index $a$ of the \cgc\
${\cal C}^{\la\lap;\lapp,a}_{\lambda\lambda'}$ runs over the values
$1,2,\ldots,\nlll$.

In the present context I am interested in an interpretation of the
Kronecker product which puts a slightly different emphasis than
the category theoretic point of view. Namely,
consider the direct sum on the right hand side
of (\ref a) as a formal sum of the objects \lla, and the tensor
product as a formal product ``\,$\star$\," on these objects; then
the set of modules \lla\ spans a ring over the integers, with
various additional properties corresponding to the respective
properties of the tensor product decomposition. The ring structure
obtained this way is an interesting object in its own right,
irrespective of the particular realization in terms of (\ref a).
Consequently one introduces an abstract ring by formalizing (some of)
these properties:
\begin{quote}{\sc Definition:} A {\em fusion ring\/} is a ring over
the integers \zet, such that the following axioms are fulfilled:
\begin{itemize} \item[~({\bf F1})~] Commutativity.
\item[~({\bf F2})~] Associativity.
\item[~({\bf F3})~] Positivity: Existence of a basis
with non-negative structure constants.
\item[~({\bf F4})~] Conjugation: Existence of an element of the
basis required by (F3) such that the evaluation of the product
with respect to this basis element furnishes an involutive automorphism.
\end{itemize}\end{quote}
 As a set, such a ring $\cal A$ is isomorphic to the lattice
$\zet^{{\rm dim}_\zet{\cal A}}_{}$. To make the properties (F1) to
(F4) more transparent, consider a basis $\{\phi_i\,|\,i\in I\}$, with
$I$ some index set of cardinality $\Ii:={\rm dim}_\zet{\cal A}$.
Denote the structure constants in this basis by \nijk, i.e.\ write
  \be  \phi_i\star\phi_j=\sum_{k\in I}\nijk\,\phi_k \,.  \labl f
Then commutativity means
  \be  \mbox{(F1)$_{\cal N}^{}$\hspace{5 em}} \nijk=\njik,
  \mbox{\hspace{14.6 em}}  \labl g
while associativity is expressed by
  \be  \mbox{(F2)$_{\cal N}^{}$\hspace{5 em}}\sum_{k\in I}
  \nijk\nklm=\sum_{k\in I}\njlk\nikm.\mbox{\hspace{6.6em}}  \labl h
The positivity axiom (F3) states that there exists a basis $\{\phi_i\}$
such that
  \be  \mbox{(F3)$_{\cal N}^{}$\hspace{5.9em}}
  \nijk\in\zetpluso . \mbox{\hspace{14.2em}}   \labl{13}
Finally, the axiom (F4) requires that for this choice of basis there
exists an index $i_\circ\in I$ such that, first, the {\em conjugation map\/}
$\phi_i\, \mapsto\,\sumi j C_{ij}\phi_j$ with
  \be  C_{jk}:={\cal N}_{jk}^{\ \;i_\circ}  \labl{cjk}
is an involution, i.e.\ that the matrix $C$ with entries (\ref{cjk})
satisfies $C^2=\one$; owing to (\ref{13}) this implies that
the map is a permutation of order two and hence can be written as
  \be  \phi_i\, \mapsto\, \phi_i^+\equiv\phi_{i^+}:=\sumi j C_{ij}\phi_j
  \ee
for some $i^+\in I$, with $(i^+)^{+_{}}_{}=i$, or in other words, that
  \be  \mbox{(F4)$_{\cal N}^{\scriptscriptstyle(1)}$\hspace{4.6em}}
  {\cal N}_{jk}^{\ \;i_\circ}\equiv C_{jk}^{}=\delta_{j,k^+}\,;
  \mbox{\hspace{11em}}   \ee
and second, (F4) requires that the conjugation map is an automorphism,
i.e.\ satisfies $\phi_{i^+}^{}\star \phi_{j^+}^{} =\sum_i\nijk
\phi_{k^+}^{} $, i.e.
  \be  \mbox{(F4)$_{\cal N}^{\scriptscriptstyle(2)}$\hspace{4.5em}}
  {\cal N}_{i^+j^+}^{\ \ \;\;k^+}=\nijk. \mbox{\hspace{13em}}   \ee
The generator $\phi_{i^+}^{}$ is called the element {\em conjugate\/} to
$\phi_i$.
The matrix $C$ is called the conjugation matrix; it can be used as
a `metric' which
lowers or raises indices. In particular one can define structure
constants with lower indices only,
  \be  {\cal N}_{ijl}:=\sum_{k\in I}\nijk C_{kl}
  = {\cal N}_{ij}^{\;\;l^+}.  \labl n
Then ${\cal N}_{ijk}\in\zetpluso$, and it follows from associativity
and commutativity
that the ${\cal N}_{ijk}$ are completely symmetric in the indices
$i,\,j,\,k$; e.g.
${\cal N}_{ikj}=\sumi l {\cal N}_{ik}^{\ \;l}{\cal N}_{jl}^{\ \;i_\circ}
=\sumi l{\cal N}_{ij}^{\ \;l}{\cal N}_{lk}^{\ \;i_\circ}={\cal N}_{ijk}$.
As a consequence of this total symmetry and of the automorphism
property of $C$, one has
  \be  {\cal N}_{i_\circ j}^{\ \ \,k} ={\cal N}_{i_\circ\, j \,k^+}^{}
  ={\cal N}_{j \,k^+\,i_\circ}^{} ={\cal N}_{j\,k^+}^{\ \ \,i_\circ^+}
  ={\cal N}_{j^+\,k}^{\ \ \ i_\circ}=C_{j^+k}^{}=\delta_{jk}^{}. \labl-
This means that the generator $\phi_{i_\circ}^{}$ is a (right, and
analogously a left) unit, so that one writes $\phi_{i_\circ}=\bfe$;
in basis independent notation, one has
$\bfe\star\phi=\phi=\phi\star\bfe$ for all elements $\phi$ of the ring.

Thus \ca\ is a unital ring, with the unit belonging to the
preferred basis which has non-negative structure constants. Moreover,
according to (\ref-) one has $C_{i_\circ j}^{}={\cal N}_{i_\circ j}
^{\ \ \;i_\circ}=\delta_{j,i_\circ}^{}$, i.e., as should be expected, the
unit is self-conjugate, or in other words, the conjugation
is a unital automorphism.

In the following, I will refer to any basis ${\cal B}_{\rm can}\equiv
\{\phi_i\mid i\in I\}$ that is singled out by the above properties
as a {\em canonical basis\/} of the fusion ring; in terms of
the usual euclidean scalar product on $\zet^\II\subset\reals^\II$,
the elements of this basis are points of length one. Also, the symbol
\nijk\ will always refer to the structure constants in a canonical
basis ${\cal B}_{\rm can}$; these structure constants are called
{\em \frc s.} Similarly, unless stated otherwise, the symbol $\phi_i$
will refer to an element of ${\cal B}_{\rm can}$.
Finally, it is conventional to use the symbol `\,0\,'
in place of $i_\circ$; thus e.g.\ the fact that \bfe\ is a unit
reads in terms of the \frc s
  \be  {\cal N}_{i0}^{\ \;j}=\delta_{ij}^{}. \ee

In the context of \findim\ \g-\rep s, it is natural to think of
only integral linear combinations of the generators $\phi_i$.
In contrast, in the application to \cft, it often
turns out to be more convenient to allow for arbitrary complex
linear combinations. Correspondingly one views the \furu s
as an algebra over \complex:
\begin{quote}{\sc Definition:} A {\em \fra\/} \fa\ \,is an \alg\ over
the complex numbers \complex, such that the axioms
(F1) to (F4) are fulfilled, with (F3) refined to
\begin{itemize} \item[~({\bf F3})$'\,$] Existence of a basis
with non-negative integral structure constants.\end{itemize} \end{quote}
If the index set $I$ is finite, \fra s acquire a few
particularly interesting properties. Therefore one defines:
\begin{quote}{\sc Definition:} A {\em \rfa\/} \fa\ \,is a \fra\
for which the following axiom holds:
\begin{itemize} \item[~({\bf R})\,~] Finite-dimensionality.
\end{itemize} \end{quote}
Note that the \fra s furnished by the tensor products (\ref a) are not
rational (by property (c), they are however quasi-rational in the sense that
for fixed $i$ and $j$, $\sum_{k\in I}\nijk$ is still finite).
In contrast, in the sequel the main interest will be in rational
\fra s, and the property (R) will be assumed unless stated
otherwise. Sometimes it is possible to construct from a non-rational
\fra\ an associated rational one by a certain process of `truncation';
examples of this phenomenon arise in the \rep\ theory of \qg s
with deformation parameter a root of unity \cite{Algs,pasa,fugp2,fuva2},
and in the context of \wzwts\ that will be the subject of section 5.
In the category theoretical setting, such a truncation implies
that the commutativity isomorphisms can no longer be
involutive, so that the category supplied by the \furu s
is a rigid {\em quasitensor\/} (or {\em braided monoidal\/}) category.
As a consequence, generically the tensor category is not
tannakian, i.e.\ there does not
exist a compatible tensor functor from this category
to the category of \findim\ vector spaces; this is
in contrast to, for example, the non-rational \fra s that are supplied
by \g-tensor products, where such a functor is provided by the
forgetful functor.

It is clear that two \fra s must be considered as equivalent
if they have the same dimension and if there exists a bijection $\sigma$
of the index sets $I^{(1)}$ and $I^{(2)}$ which is compatible with
(F1) to (F4), i.e.\ is such that for $\phi_{\sigma(i)}^{(1)}\,\hat=\;
\phi_i^{(2)}$ one has $\phi_0^{(1)}\,\hat=\; \phi_0^{(2)}$,
and $\phi_{\sigma(i^+)}^{(1)}=\phi_{(\sigma(i))^+}^{(1)}$, as well as
$({\cal N}^{(1)})_{\sigma(i)\,\sigma(j)}^{\ \ \ \ \ \ \sigma(k)}=
({\cal N}^{(2)})_{ij}^{\ \;k}$. This equivalence relation will
be called isomorphism, and the map $\sigma$ a unital $^*$-isomorphism
of the \fra s, or shortly a \furu\ isomorphism. Similarly, an endomorphism
of a \fra\ with analogous properties is called a {\em \furu\ automorphism}.

It should be pointed out that the axioms of a fusion ring (or a \fra)
do not require the canonical basis to be uniquely determined. For instance,
consider the \twodim\ fusion ring with basis $\{\bfe,\phi\}$ and
products $\bfe\star \bfe=\bfe$, $\bfe\star \phi=\phi=\phi\star \bfe$,
and
  \be  \phi\star \phi=\bfe+\phi \,.  \labl{ff}
Clearly, this basis is canonical. But owing to
  \be  (\bfe-\phi)\star(\bfe-\phi)=\bfe-2\,\phi+\phi\star\phi=
  \bfe+(\bfe-\phi),  \ee
the mapping $\bfe\mapsto\bfe$, $\phi\mapsto\bfe-\phi$ defines a \furu\
automorphism, and hence in particular $\{\bfe,\bfe-\phi\}$ is a canonical
basis as well. Still, the freedom that remains when choosing a canonical
basis seems rather insignificant, so that in spite of this non-uniqueness
I will stick to the qualification `canonical.'

To conclude this introduction into the subject, let me mention that there
is a nice pictorial \rep\ of \frc s and their properties. Describe
the \frc s by labelled trivalent graphs with ordered oriented lines,
according to
  \be  \begin{array}{ccc} \mbox{}\\[-9.8 mm] \nijk=&\smallpicnijk ijk
  & . \\[4 mm] {} \end{array} \ee
Then the conjugation matrix corresponds to
  \be  \begin{array}{ccccc} \mbox{}\\[-11 mm] C_{ij}=&\smallpicnijo ij0
  & \cong & \picij & , \\[4 mm] {} \end{array} \ee
so that the \frc s with three lower indices are represented by graphs
with three incoming lines,
  \be  \begin{array}{ccc} \mbox{}\\[-9.8 mm] {\cal N}_{ijk} =&
  \smallpicnijuk ijk & , \\[7 mm] {} \end{array} \ee
while associativity together with commutativity result in the
{\em duality}
  \be  \begin{array}{ccccccc} \mbox{}\\[-14 mm] \dsumi m &
  \smallpicschannel jimkl  & = \dsumi n &
  \smallpictchannel jinkl  & = \dsumi p &
  \smallpicuchannel jkpil & \mbox{\hspace{-9 mm}}. \\[9 mm] {}
  \end{array} \labl G

The remainder of this paper is organized as follows.
In section 2 the diagonalization of the \furu s is described,
and the notion of modular \fra s is introduced. Section 3
provides a short introduction to those aspects of \twodim\ \cft\
that are connected with the concept of \furu s.
In section 4 the relevance of the modular group to \fra s
and to \cft\ is explained. Section 5 provides a brief
description of a particular class of \cfts, namely \wzwts,
some of which are used as examples in the succeeding sections.
In section 6 fusion rings are analyzed in terms of polynomial rings.
In particular the possibility that the ideal that has to be divided
out from a free polynomial ring in order to obtain the fusion ring
derives from a potential is investigated in some detail.
To provide examples for some of the concepts introduced in sections
2 to 6, various aspects of the \furu s of the \sln\ \wzwt\
at arbitrary level are discussed in section 7.
In section 8 I explain the concept of fusion graphs and describe
the graphs obtained in various special cases.
Section 9 contains an overview of what is known about the
classification of (modular) \fra s; in particular the list of
polynomial \fra s with generator of fusion dimension $\leq2$ is
presented.
I conclude in section 10  with further remarks on the connection
between the fusion rules and the \opa.

In short, there are a few sections dealing mainly with \cft\ aspects
(sections 3, 5, 10), while in the other sections \furu s are mainly
considered as abstract rings or
algebras. There are of course various interrelations among the
different aspects, but some sections may be read without the need to
know the contents of all of the preceding sections. The logical
interdependence is essentially as depicted in the following diagram:
  $$  \begin{array}{cccccccccc}
  1&\rightarrow& 2&\rightarrow& 8&\rightarrow& 9\\[2.1 mm]
  &&&\searrow &&\nearrow\\[2.1 mm]
  && 3&\rightarrow& 4&\rightarrow& 5\\[2.1 mm]
  &&&\searrow &&\searrow &&\searrow\\[2.1 mm]
  &&&& 10&& 6&\rightarrow& 7&.  \end{array} 
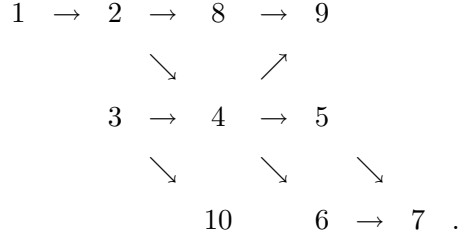
$$

\subsect{Diagonalization}

The structure constants of a commutative associative algebra play
the role of the
\rep\ matrices of the algebra in its regular (or adjoint) \rep.
Accordingly, for a \fra\ with structure constants \nijk, the {\em
fusion matrices\/} \cn i, $i\in I$, defined as the matrices with entries
  \be  (\cn i)_{jk}=\nijk, \ee
furnish a \rep\
  \be  \pireg: \ \ \phi_i\,\mapsto\,\pireg(\phi_i) :=\cn i  \ee
of the algebra. From the fact that the
conjugation $C$ is an automorphism, one learns that $\cn{i^+}=
(\cn i)_{}^t$, while associativity and commutativity of the \furu s
imply that the fusion matrices commute among another. In particular,
$[\cn i,(\cn i)_{}^t]=[\cn i,\cn{i^+}]=0$, i.e.\ the fusion matrices
are normal.

Since a \fra\ is abelian, all its
irreducible \rep s are \onedim. Moreover, because of their normality
the \cn i are diagonalizable by a unitary matrix,
in such a way that the orthogonal eigenvectors
$v_j$ and hence the diagonalizing matrix do not depend on the
index $i$ of \cn i. In other words,
the regular \rep\ of the \fra\ is fully reducible, and
is hence the direct sum of \Ii\ inequivalent \onedim\ \rep s.
I will denote the diagonalizing matrix by \dm, but (in order
to allow for a convenient normalization of the eigenvectors) do not
require that \dm\ be unitary.
Thus $(\dm^{-1}\cn i\dm)_{jk}=\nv ji\delta_{jk}$, and
the eigenvalues \nv li of the fusion matrices
furnish \onedim\ \rep s of the algebra, i.e.
  \be  \sumi k \nijk \nv lk=\nv li \nv lj . \labl E
Note that, being solutions to the characteristic equation ${\rm det}
(\cn j-\nv ij\one)=0$, the eigenvalues \nv ij are \alg ic integers.

Comparing (\ref E) with the eigenvector equation
$\sumi k (\cn i)_{jk}\vv lk=$ $\nv li \vv lj,$
one learns that the eigenvalues coincide with the (appropriately
normalized) eigenvectors,
  \be  \dM{ji}\equiv \vv ij = \nv ij .  \labl{23}
In other words, there exists a set of $|I|$ vectors $v_i$ such that
  \be  \pi_j: \ \ \phi_i\,\mapsto\,\pi_j^{}(\phi_i):=\vv ji  \labl A
for $j\in I$ are \onedim\ \rep s of the \fra, satisfying
  \be  \sumi k \nijk \vv lk=\vv li \vv lj . \labl{36}
This  may also be written as
  \be  \nijk=\sumi l \dM{il}\dM{jl}(\dm^{-1})_{lk} . \labl*

A \frsa\ \cat\ of a \fra\ \ca\ is a subalgebra
$\cat\subseteq\ca$ which is itself a \fra. The unit $\tilde\phi_0$
of \cat\ coincides with the unit $\phi_0$ of \ca\ ($\tilde\phi_0$ must
belong to a canonical basis of \cat, and extending this
basis to a basis of \ca\ shows that $\tilde\phi_0$ is a unit for
all of \ca, which is however unique). Thus for $\phi\not\in\cat$,
also $\tilde\phi_0\star \phi\not\in\cat$, implying that a \fra\ does
not possess any non-trivial (unital) ideals. Hence \fra s are simple.
In particular, in the rational case, i.e.\ for $|I|<\infty$,
the algebra is a \findim\ simple associative algebra.
As a consequence \cite{CUre2,kawA}, the \onedim\ \rep s are
actually exhausted by the \rep s (\ref A); in other words,
the index set $I$ not only labels the generators, but also the
inequivalent irreducible \rep s of the algebra.

Because of the orthogonality of the eigenvectors $v_i$,
the right inverse of the matrix \dm\ has entries
  \be  (\dm^{-1})_{ij}=\eta_i^{\,2}\,\dm^*_{\!ji} \,,   \labl D
where
  \be  \eta_i^{}:=\mbox{\large(}\sumi j |\dM{ji}|^2 \mbox{\large)}
  _{}^{-1/2} .    \labl B
Thus (\ref*) can be rewritten as
  \be  \nijk=\sumi l \dM{il}\dM{jl}\dm^*_{\!\!kl} \,
  \mbox{\large(}\sumi m |\dM{ml}|^2 \mbox{\large)}
  _{}^{-1/2} .    \ee
Since the right inverse must coincide with the left inverse,
one not only has the orthogonality relation
  \be  \sumi l \dM{li}^{}\dm^*_{\!lj}=\eta_i^{-2}\delta_{ij}^{},  \ee
but also
  \be  \sumi l \eta^2_l\,\dM{il}^{}\dm^*_{\!jl}=\delta_{ij}^{}.  \ee

Since the matrix elements of a fusion matrix \cn i are non-negative,
it follows from standard results in the theory of matrices
(see e.g.\ \cite{SEne,GAnt,PEas})
that their largest eigenvalue is a positive real number (and is
non-degenerate if the matrix \cn i is indecomposable), and that there
exists a unique normalized eigenvector
to this eigenvalue with only positive entries, called the
Perron\hy Frobenius eigenvector of \cn i. I will call the
corresponding eigenvalue the {\em fusion dimension\/} and denote
it by \cd i; in other contexts, this Perron\hy Frobenius eigenvalue is
known as statistical dimension of a superselection sector
(in algebraic field theory \cite{frrs}) or of an object of a quantum
category \cite{kerl}, as quantum dimension (in \cft \cite{verl2}, and
in the theory of \qg s \cite{resh2}), or as the
square root of the index of an inclusion of von Neumann algebras
\cite{jone,long3}.
As noticed after (\ref E), the fusion dimension is
an algebraic integer. But actually not any
positive algebraic integer qualifies as a fusion dimension; for
instance (see e.g.\ \cite{GOdj,FRke,jf17}), it follows
from an old result by Kronecker \cite{kron} that
  \be  \cd i\in\{2\,{\rm cos}(\frac\pi n) \mid n\in\zet_{\geq3}\}
  \cup[2,\infty). \labl H
Because of $\cn{i^+}= (\cn i)_{}^t$, the matrices \cn i and \cn{i^+}
have complex conjugate eigenvalues,
  \be  (\vv li)^*=\vv l{i^+}.  \labl{311}
In particular, $ \cd i = \cd{i^+}$.
Since \dm\ diagonalizes all of the matrices \cn i, they have a unique common
Perron\hy Frobe\-nius eigenvector. Also, one may order the
eigenvalues such that this is the vector $v_j$ with $j=0$, so that
  \be  \cd i=\vv 0i=\dM{i0} \,.  \labl{j=0}
On the other hand, the fusion matrix \cn 0 for the identity $\phi_0$
is just the unit matrix, so that
  \be  \dM{0i}=1   \labl{313}
for all $i\in I$.

{}The diagonalizability implies among other things
that apart from the basis ${\cal B}_{\rm can}$ of fields $\phi_i$
there exists another distinguished basis, whose elements are given by
  \be  e_i:=\eta_i^2\sumi j \dm_{\!\!ji}^*\,\phi_j  .  \labl{ef}
Because of the relation (\ref D), the inverse basis transformation reads
$\phi_i=\sumi j \dM{ij}e_j.$ By applying first (\ref{ef}),
(\ref f), and the inverse basis transformation, and then (\ref*),
to the fusion product $e_i\star e_j$, one finds that
  \be  e_i\star e_j=\delta_{ij}\,e_j  ; \ee
thus the elements $e_i$ are primitive idempotents
(the existence of a basis of primitive idempotents
is guaranteed \cite{CUre2} because \fra s are simple unital
associative \alg s). In other words, in the basis $\{e_i\}$
the \furu s are diagonal, i.e.\ the structure constants read
$\tilde{\cal N}_{ij}^{\ \;k}=\delta_{ij}\delta_{ik}$, and the
components of any element of \ca\ in this basis are just its
eigenvalues in the regular \rep. Further, as follows with the help of
(\ref{313}), the $e_i$ provide a partition of the unit element,
  \be  \phi_0=\sumi j e_j. \ee

For the subsequent discussion it will be convenient to pass to a
normalization of eigenvectors different from the one used so far. Thus
consider the matrix
  \be  \yy ij:=\eta\,\cd j\,\dM{ij},   \labl{dm}
with $\eta=\yy00$ some positive real number to be fixed in (\ref C)
below.  Note that $\yy i0=\eta\cd i$ obeys $ \yy i0=\yy 0i=\dy_{\!i0}^*$.
Let me make in the sequel the {\em non-trivial assumption\/} that this extends
to a symmetry of the whole matrix \dy; this may be reformulated as
the following axiom.
\begin{quote} \begin{itemize} \item[~({\bf M1})~] Symmetry of $Y$:
\end{itemize} \end{quote}
  \be  \cd i\,\vv ij=\cd j\,\vv ji.  \labl Y
Let me point out that without this symmetry requirement, the labelling
of the eigenvectors $v_i$ and that of the matrices \cn i are logically
completely independent. In contrast, by imposing (M1) one correlates
these two types of labelling. In other words, one defines a bijection
(compatible with the choice (\ref{j=0})) between the \rep s $\pi_i$
and the represented objects $\phi_i$. (Incidentally, this also implies
a partial fixing of the freedom that remains in the choice of the
canonical basis.)

(M1) is also equivalent to the requirement that
the numbers $\eta_i$ defined in (\ref B) must be
proportional to the fusion dimensions \cd i. Also, along with (\ref{311})
it follows immediately that $\yy i{\,j^+}=\dy_{\!ij}^*=\yy{i^+\,}j$, i.e.
  \be  \dy C =\dy^*=C\dy . \labl{yc}
Moreover, since \dm\ is invertible and diagonalizes the \furu s, \dy\ is
invertible, too, and
 $ \sumi k \nijk\,\yy kl=\yy il \yy jl/ \yy 0l,$ as well as
  \be  \sumi{j,k} \yz mj\nijk\,\yy kl=(\yy im/ \yy 0m)\,\delta_{ml}.
  \labl{317}
The latter result may be transformed into an expression for the \frc s
analogous to (\ref*),
  \be  \nijk=\sumi l \frac{\yy il \yy jl \yz kl}{\yy 0l} \,. \labl{df}
The inverse of \dy\ is related to the inverse of \dm, $\yz ij=(V^{-1})^{}
_{ij}/\eta \cd i$, which by (\ref D) implies that it is proportional
to $\dy^*$. Thus without loss of generality one may impose that
\dy\ be unitary, which amounts to requiring
  \be  \eta_j^{}=\eta\,\cd j    \labl{ed}
for all $j\in I$, thereby fixing in particular the constant $\eta$ as
  \be  \yy00\equiv\eta=\eta_0= \mbox{\large(}\sumi j \cd j^{\!2}
  \mbox{\large)}_{}^{-1/2} .    \labl C
Due to $\yy i0=\yy 00\cd i$ and $\cd i\geq1$, one thus has
  \be  \yy i0\geq\yy00>0. \ee
Also, together with (\ref{yc}) it follows that
  \be  \dy^2=C,  \labl{2c}
and that (\ref{df}) may be rewritten more symmetrically as a formula for
the \frc s with three lower indices,
  \be  {\cal N}_{ijk}^{}=\sumi l \frac{\yy il \yy jl \yy kl}{\yy 0l}. \ee

Let me stress that the constraint (\ref Y) is very selective;
for a {\em generic\/} \fra, \dy\ cannot be chosen symmetric. As a simple
counter example, consider the three-dimensional \fra\ with fusion
matrices
  \be  \cn0=\one, \quad \cn1=\matrx{ccc} {0&1&0\\[1 mm]1&0&0\\[1 mm]
  0&0&1} , \quad \cn2=\matrx{ccc} {0&0&1\\[1 mm]0&0&1\\[1 mm]1&1&1}.\ee
These are diagonalized by the unitary matrix
  \be  \dm=\frac1{\sqrt6}\matrx{ccc}{1&\sqrt2&\sqrt3 \\[1 mm]
  1&\sqrt2&-\sqrt3 \\[1 mm] 2&-\sqrt2&0}  \ee
(namely, $\dm^{-1}\cn1\dm={\rm diag}(1,1,-1),\
\dm^{-1}\cn2\dm={\rm diag}(2,-1,0)$), but \dm\ is not symmetric.
(More general constructions of \fra s with non-symmetric unitary
diagonalization matrix can be found in \cite{eliz}.)

Rather than delving into the general case of non-symmetric unitary
diagonalization matrix, I will turn my attention to a class of \fra s
that is even more restricted.  Namely, let me require that not only \dy\
is symmetric (i.e.\ property (M1) holds), but in addition that there exists a
diagonal \Ii$\times$\Ii\ matrix $T$ with the following properties:
\begin{quote} \begin{itemize} \item[~({\bf M2})~] $T$-matrix:
$\,T={\rm diag}(t_i)$ satisfies
\end{itemize} \end{quote}
  \be   T^*=T^{-1}, \qquad CT=T, \labl{ct}
\mbox{}\hskip4.7em and
  \be  YTY=T^*YT^*. \labl{yt}
Note that the 00-component of the latter relation reads
  \be  \sumi j t_j \cd j^2=(t_0^*)^2_{}/\yy00 .  \labl w

For reasons to be clarified in section 4 below, I will refer to the
properties (M1) and (M2) as {\em modularity constraints\/} and
to (rational) \fra s satisfying them as {\em modular \fra s}.
For modular \fra s, one finds that \dy\ can be expressed as
  \be  \yy ij=t_0^*t_i^{}t_j^{}\yy00\sumi k{\cal N}_{ik}^{\ \;j}
  t_k^*\cd k  ,  \labl{ty}
as may be checked by first applying (\ref{df}), and then twice
(\ref{yt}), to the right hand side.
Furthermore,
  \be  t_j\cd j=t_0^*\sumi k t_k^* \cd k \yy jk,  \labl{tz}
as can be seen by combining the right hand side with (\ref{ty}),
then using the \rep\ property of the fusion dimensions, and then
applying (\ref w).
Also \cite{dego}, for a modular \fra\ the \furu\ eigenvalues are
not just algebraic integers (subject to the constraint (\ref H)),
but they belong to the algebraic integers of a cyclotomic
extension of {\dl Q}.

As it turns out, the diagonalization matrix of a modular \fra\
is essentially unique. Of course,
\dy\ is not at all determined uniquely by the
requirement that it diagonalizes the \furu s. But requiring the
matrix to be also symmetric and unitary, and to obey  $\dy^2=C$ and
$\yy i0\geq\yy00>0$, determines it up to permutations in the
following sense \cite{degi}: if $\dy^{(1)}$ and $\dy^{(2)}$ are two
matrices sharing all these properties, then there exists a \furu\
automorphism $\sigma$ such that $\dy^{(1)}_{\!ij} = \dy^{(2)}_
{\!\sigma(i),j} = \dy^{(2)}_{\!i,\sigma(j)}$.
 Imposing also the property (\ref{yt}), it follows \cite{jf19}
that this automorphism must be of order three and commute with
$T$, i.e.\ $t_{\sigma(j)}=t_j$. Thus one may recover the
matrix \dy\ from the \furu s up to at most a minor
ambiguity, and in fact an efficient algorithm
(employing standard routines for the numerical diagonalization
of matrices) for doing so is available \cite{fuva3,jf19}.

In particular, the eigenvectors are essentially unique up to
scalar multiplication. This implies that any modular
\fra\ \ca\ contains an element $x\in\ca$ such that all eigenvalues
of \pix\ are distinct (if this were not so, then
the freedom in defining the eigenvectors corresponding to
degenerate eigenvalues would be a non-trivial unitary matrix).

\subsect{Conformal field theory}

Let me now describe how \fra s emerge in \cft.  To start, it is
appropriate to recall a few basic facts of \cft, or more precisely, of
the bootstrap approach to \twodim\ \cft.  The two main ingredients are
conformal invariance and the bootstrap idea.  Conformal invariance
\cite{poly2,scsw,luma,kuRy} of a \twodim\ field theory implies
\cite{luma2,bepz} that the collection of all (properly interpreted)
`fields' of the theory carries
a \rep\ of the Virasoro \alg\ $\cal V$, which is the complex Lie
\alg\ with basis $\{C\}\cup\{L_m\mid m\in\zet\}$ and Lie brackets
  \be  [L_m,L_n]=(m-n)\,L_{m+n}+\frac1{12}\,(m^3-m)\,\delta_{m+n,0}\,C\,,
  \qquad [L_m,C]=0.  \ee
More precisely, the fields $\varphi(z,\bar z)$ can be organized into a
direct sum of \ihwm s of $\cal V$ which have a common eigenvalue of
the central generator $C$. This eigenvalue $c$ is called the conformal
central charge of the theory. Fields corresponding to \hw\ vectors
are referred to as primary fields and generically denoted by $\phi$,
while those corresponding to non-\hw\ vectors are called descendant
fields.  Thus primary fields obey
  \be  {\cal U}({\cal V}_+)\,\phi=0, \qquad
  {\cal U}({\cal V}_0)\,\phi=\complex\phi, \ee
while descendants are of the form
  \be  \varphi\in {\cal U}({\cal V}_-)\,\phi, \ee
and the complete collection of descendants is obtained from the
Verma module defined by these equations by taking the unique
irreducible quotient. Here ${\cal V}_\pm$ denote the Lie sub\alg s
spanned by the generators $L_n$
with $n>0$ and $n<0$, respectively, and ${\cal V}_0$ the subalgebra
spanned by $L_0$ and $C$, while ${\cal U}({\cal L})$
stands for the universal enveloping algebra of a Lie algebra $\cal L$.

The second ingredient, the bootstrap hypothesis, amounts to
the requirement that upon forming radially ordered
products the fields of the theory constitute a closed associative
operator algebra \cite{kada,wils,poly3,scsv},
  \be  \varphi(z,\bar z)\,\varphi'(w,\bar w)=\sum_{\varphi''}
  C_{\varphi\varphi'}^{\ \ \varphi''}(z,w,\bar z,\bar w)\,
  \varphi''(w,\bar w).   \labl{34}
To be precise, this must be valid inside all correlation functions, which
are defined (via the correspondence \cite{bepz} between fields $\varphi$
and state vectors in the physical Hilbert space of the theory) as the
vacuum expectation values $\langle\, \ldots \varphi(z,\bar
z)\,\varphi'(w,\bar w)\ldots\,\rangle\,$ of radially ordered products of
fields.  The radius of convergence of the expansion (\ref{34}) is then
determined by the remaining fields that are present in the correlator.

In the above, the variables $z$ and $\bar z$ are complex coordinates
for the \twodim\ space-time on which the theory is living; for many
aspects of the theory, these variables
need not be considered as complex conjugates
of each other, but can be treated as independent.  The identity primary
field \bfe, which by definition satisfies $\bfe\,\varphi(z,\bar
z)=\varphi(z,\bar z)$ for all fields $\varphi$ of the theory, does not
depend on these variables at all. The generating function
$T(z):=\sum_{m\in\zet}L_m\,z^{-m-2}$ plays the role of the \emt\ of
the theory and is a descendant of the identity primary field.
If there are further fields $W(z)$
which do not depend on the antiholomorphic coordinate $\bar z$, then it
is convenient (and, for some aspects, even mandatory) to consider instead
of the Virasoro algebra a larger algebra \w\ containing $\cal V$ as a Lie
subalgebra. Similarly as the Virasoro generators $L_n$ are the Laurent
modes of $T(z)$, the additional generators of \w\ are the Laurent modes of
the fields $W(z)$. Furthermore, treating the antiholomorphic coordinate
$\bar z$ in an analogous manner, one arrives at a direct sum
$\w\oplus\overline{\cal W}$, with $\overline{\cal W}$ generated by the
modes of purely antiholomorphic fields $\overline W(\bar z)$.  The algebra
$\w\oplus\overline{\cal W}$ is called the symmetry algebra of the theory,
and its summands the (holomorphic, respectively antiholomorphic) chiral
\alg s of the theory. If {\em all\/} fields which depend either only
on $z$ or only on $\bar z$ are included into the symmetry algebra, then
it is called maximally extended or, shortly, maximal.

The remarks made above about the role of the
Virasoro algebra extend to analogous statements about the symmetry
algebra.  Accordingly, in the sequel I will label the primaries of a
given \cft\ by an index set $I$, and denote by $[\phi_i]$, $i\in I$, the
collection of fields which correspond to the \ihwm\ of \w\ whose \hw\ is
carried by the primary $\phi_i$; I will call $[\phi_i]$ the
\w-{\em family\/} of $\phi_i$.
The algebra \w\ is endowed with a \zet-gradation supplied by the
mode numbers $n$ of the non-central generators $W_n$, and $L_0$
acts as a derivation,
  \be  [L_0,W_n]=-n\,W_n .  \labl{zg}

The requirement of conformal, respectively \w-, symmetry imposes severe
constraints on operator products.  In particular, operator products of
primaries can be written as
  \be  \phi_i(z,\bar z)\,\phi_j(w,\bar w)=\sumi k
  C_{ij}^{\ \,k}\,(z-w)^{-\Delta_i-\Delta_j+\Delta_k}_{}
  (\bar z-\bar w)^{-\bar\Delta_i-\bar\Delta_j+\bar\Delta_k}_{}
  \,\left[ \phi_k(w,\bar w) + \ldots\, \right] .  \labl o
Here $\Delta_i,\;\bar\Delta_i$ are the conformal dimensions of $\phi_i$,
i.e.\ eigenvalues of $L_0$ and $\bar L_0$, the ellipsis stands for
terms involving descendant fields, and $C_{ij}^{\ \,k}$ are complex
numbers, known as \opc s.  By applying elements of $\,{\cal U}(\w)$ to
both sides of (\ref o), one obtains the so-called \w-Ward identities
which relate the \opc s involving different members of a given \w-family.
As a consequence, one can extract the basis independent content of the
\opa\ in a manner analogous as one gets the decomposition (\ref a) of
tensor products of \rep s of a simple Lie algebra from the Clebsch\hy
Gordan series (\ref e).  This way one arrives at the \furu s \cite{verl2}
of a \cft; they are written as in the formula
(\ref f), where \nijk\ describes now
the number of distinct couplings among the \w-families headed by the
primary fields $\phi_i,\ \phi_j$, and $\phi_k$ that occur in the
operator product (\ref o).

The notion of fusion rings and \fra s as introduced in section 1 has
in fact been tailored to fit to the \furu s arising, in the manner
just described, in
\cft. Namely, the canonical basis required by the axiom (F3) is
supplied by the primary fields $\phi_i$, the commutativity (F1)
follows from the fact that the families $[\phi_i]$ and $[\phi_j$]
appear symmetrically in the definition of \nijk, and the associativity
(F2) is a consequence of the associativity of the \opa. Finally,
the unit $\phi_0$ is provided by the identity primary field \bfe, and
the field conjugate to $\phi_i$ whose existence is required by (F4) is
the primary field $\phi_i^+$ that satisfies
  \be  \langle\phi_i^{}(z,\bar z)\,\phi_i^+(w,\bar w)\rangle=
  (z-w)^{-2\Delta_i} (\bar z-\bar w)^{-2\bar\Delta_i}   \labl{37}
(this conjugate field exists and is unique, as a consequence of the
fact that any one-point function $\langle\phi\rangle$ vanishes except
for $\phi=\bfe$, while not all two-point functions
$\langle\phi\varphi\rangle$ can vanish \cite{bepz}). Furthermore,
there are large classes of \cfts\ which have a \fra\
satisfying also the axiom (R), which means that the number of primary
fields is finite. This possibility arises as a result of the presence
of null vectors in the Verma modules of \w\ (these lead to the
decoupling of \w-families which naively would be expected to
contribute to the \opa). It is commonly supposed that the
collection of all \twodim\ \cfts\ can be endowed with a topology
such that the rational theories constitute a dense subspace.

Let me mention that
one has to be rather careful if one wants to read off
 \footnote{~In practice, one rarely proceeds in this direction, since
there there does not exist a simple algorithm for computing operator
products without using the \furu s as an input.}
 the \frc s from the \opa.
This is so because coefficients $\nijk>1$ are allowed, as typically
happens if the zero mode (or `horizontal') subalgebra
\wo\ of \w\ is nonabelian.
 In this situation the \opc s as introduced above are actually not
complex numbers, but rather complex numbers
multiplied by appropriate invariant tensors of \wo. For a given
triple of \w-families there may exist several independent such tensors
$\eta$; accordingly, $C_{ij}^{\ \,k}$ gets replaced by
$C_{ij}^{\ \,k;(1)}\eta_{ij}^{\ \,k;(1)}+
C_{ij}^{\ \,k;(2)}\eta_{ij}^{\ \,k;(2)}+\ldots\;$.
(incidentally, the tensors $\eta_{ij}^{\ \,k;(a)}$ are always
such that, with a suitable labelling of the degeneracy index $a$,
they vanish for the grades $\,0,\,1,\,\ldots,a-2$; in particular,
there is a unique coupling among the {\em primaries\/} $\phi_i,\
\phi_j$ and $\phi_k$, and when raising the grade by one unit,
at most one new coupling arises \cite{jf17}).

Note that the \frc s as introduced above for a \cft\ refer only to the
holomorphic symmetry algebra \w, and not to its antiholomorphic
counterpart $\overline{\cal W}$.  This is consistent provided that the
maximal symmetry algebra is chosen; namely, it can be shown
\cite{diVe,mose3} that in this situation there exists a (possibly
trivial) permutation of the index set $I$ which furnishes an
isomorphism between the holomorphic and the antiholomorphic \fra s.

Let me also mention another aspect of the \opa.
{}From (\ref o) and (\ref{37}) it follows that the \opc s can be read off
the three-point functions of primary fields as $\cijk=\lim_{z,\bar z
\rightarrow\infty}$ $z^{-2\Delta_k}\bar z^{-2\bar\Delta_k}
\langle\phi_i(0,0)\phi_j(1,1)\phi_k(z,\bar z) \rangle$, and in fact the
three-point functions are, up to their normalization \cijk, uniquely
fixed by the \w-Ward identities.  However, except for the case of free
field theories it is not possible to read off these normalizations as
well. But what one can often do, is to compute all \four s
   \be  \calfzz\equiv {\cal F}_{ijkl}(z,\bar z)
  =\langle\phi_i(z,\bar z)\phi_j(0,0)\phi_k(1,1)\phi_l
  (\infty,\infty)\rangle ,   \labl{11}
albeit again only up to normalization, and factorize them
into three-point functions. The normalizations can then be determined
(up to the freedom given by the normalization of the
primary fields themselves) with the help of the associativity
of the \opa.

An important ingredient in the calculation of correlators such as
(\ref{11}) is the concept of chiral blocks, which arises naturally
from a few basic properties of \cfts. Namely, first, the vacuum vector is
invariant under
the subalgebra of $\cal V$ that is spanned by $L_0$, $L_1$ and $L_{-1}$;
this allows to put the position of three of the primaries in the
correlator to preferred values, say 0, 1, $\infty$, as has already been
done in the formula (\ref{11}). Next, the closure of the \opa\ implies that
\calfzz~can be written as a sum of products of \threepf s, and because of
\w-invariance the contributions from all fields in a fixed
\w-family $[\phi_m]$ sum up to a function
of definite analytic behaviour. The fact that the symmetry algebra is
the direct sum of two chiral halves allows to separate the $z$- and $\bar
z$-dependence of these functions. Thus the correlator becomes a sum of
products of purely holomorphic or antiholomorphic pieces, the {\em chiral
blocks}, according to
  \be  \calfzz=\sum_{m=1}^M \sum_{\bar m=1}^{\bar M} a_{m\bar m}
  {\cal F}_m(z)\bar{\cal F}_{\bar m}(\bar z). \labl{14}
The number $M\equiv M_{ijkl}$ of blocks is determined through the \frc s
as
  \be  M=\sumi n {\cal N}_{ij}^{\ \;n} {\cal N}_{nkl}. \labl M
The same expression is valid for the integer $\bar M$ (here it is implicit
that the chiral algebra is maximally extended), and
one has actually $a_{m\bar m}=a_m\delta_{\bar m,\sigma
(m)}$ for some permutation $\sigma$. If the chiral blocks are properly
normalized, the coefficients $a_m$ coincide with
the product $C_{ij}^{\ \,m}C_{mkl}$ of \opc s \cite{bepz,knza}.

The associativity of the \opa\ implies certain identities, known as
duality relations, for the system of chiral blocks of a theory.
These lead in particular to the so-called polynomial equations for
the fusing and braiding matrices, which play an important role
in the classification of \cfts\ \cite{mose,mose3}. (For additional
information, see section 10.)

To conclude this section, let me point out that the fusion product
defined in the manner described above does {\em not\/} provide a
description of the tensor products of \hwm s of the algebra \w.
For the latter, the eigenvalues of central operators such as
$C$ add up, whereas all the \hwm s appearing in a given \cft\
have the same eigenvalues. Nevertheless it is possible to think of
the collection of \w-modules of a \cft\ as the objects of a rigid
braided monoidal
category, with the tensor product of the category identified with the
fusion rules; the commutativity and associativity constraints of the
category then coincide with the (genus zero) polynomial equations
that were introduced in
\cite{mose}. As all the (non-trivial) \w-modules are \infdim, there is
no reason to expect that one can find a compatible tensor
functor from this category to the
category of \findim\ vector spaces, and indeed (compare e.g.\
\cite{kerl,drin8}) such a functor does not exist.

\subsect{The connection with the modular group}

According to the formul\ae\ (\ref{yc}), (\ref{2c}), (\ref{ct}), and
(\ref{yt}), for the special class of (quasi)rati\-o\-nal \fra s that
satisfy the modularity constraints (M1) and (M2), the matrix $S:=\dy$
diagonalizing the fusion matrices, the conjugation matrix $C$, and some
diagonal matrix $T$ obey
    \be S^2=C=(ST)^3 \labl I
and
    \be  S=S^t,\qquad S^*=S^{-1}, \qquad T^*=T^{-1} . \labl J
These are the defining relations for a unitary matrix \rep\ of the group
\slz\ of $2\times2\,$-matrices with integral entries and determinant 1,
or what is the same, for a projective unitary matrix \rep\ of the
quotient $\pslz=\slz/\{\pm\one\}$ of \slz\ by its center $\{\pm\one\}$,
which is known as the {\em modular group}. This justifies the
qualification `modular' \fra\ that I ascribed to these structures
in section 2.

A rather different role of the modular group in \cft\ emerges as follows.
The information about the conformal dimensions of the members of a
\w-family $[\phi_i]$ can be encoded into a function $\chi$ of a complex
variable $\tau$ (well-defined for Im$(\tau)>0$) according to
  \be \chii(\tau)\equiv\chi^{}_{\phi_i}(\tau):={\rm tr}^{}_{[\phi_i]}
  {\rm e}^{2\pi\ii\tau(L_0-c/24)} \,;  \ee
$\chi_i$ is called the (Virasoro-specialized) character of the family.
Conjugate families possess identical characters,
  \be  \chi^{}_{i^+}=\chii.  \labl{ci}
The derivation property (\ref{zg}) of $L_0$ induces
a \zet-gradation on the \w-modules. More precisely, one defines the
grade of primary fields to be zero and, recursively, grade($W_{-m}
\varphi)=$\,grade($\varphi)+m$. Thus
the conformal dimension of a field $\varphi\in
[\phi_i]$ is just the sum of the conformal dimension $\Delta_i$
of $\phi_i$ and the grade of $\varphi$, so that
  \be \chii(\tau)={\rm e}^{2\pi\ii\tau(\Delta_i-c/24)}\,\mbox{$\sum
  _{n\in\zetplusO}$}d_n {\rm e}^{2\pi\ii\tau n} ,  \ee
with $d_n$ the number of descendants at grade $n$.
In particular, it follows immediately that
  \be \chii(\tau+1)={\rm e}^{2\pi\ii\tau(\Delta_i-c/24)}\,
  \chii(\tau). \labl1

Now \pslz~can abstractly be defined as the group generated freely by
two elements \sx\ and \tX\ modulo the relations
$\sx^2=(\sx\tX)^3=\id.$ These generators can be realized as the
transformations $\sx\!:\,\tau\mapsto-1/\tau,\;\,\tX\!:\,\tau\mapsto\tau+1$ of
some complex variable $\tau$ which may be restricted to the upper complex
half plane.  According to (\ref1), the transformation \tX\ acts on the
characters $\chi$ as multiplication by a diagonal matrix $T$ with entries
  \be  T_{jk}=\delta_{jk}\,t_k\equiv
  \delta_{jk}\,\exp(2\pi\ii(\Delta_j-c/24)).  \labl T
Note that together with the formul\ae\ (\ref C), (\ref w) and (\ref{tz})
this implies the relation
  \be  c=\frac4\pi\,{\rm arg}\mbox{\large(}\sumi j \cd j^2\,{\rm e}
  ^{2\pi\ii\Delta_j}\mbox{\large)} \ee
between the conformal dimensions, the fusion dimensions and the conformal
central charge, which determines $c$ modulo 8.
There are arguments, to be described in some detail further on, that
the transformation \sx\ acts on the collection of
characters as multiplication by a matrix, too. Denoting this matrix
by $S$, one arrives at the relations (\ref I) and (\ref J) above,
with the entries of $T$ specified by the expression (\ref T).
That one obtains a \rep\ of \slz, and hence generically
a {\em projective\/} \rep\ of \pslz,
is not in conflict with the identities $\sx^2={\sl id}=(\sx\tX)^3$,
because due to the equality (\ref{ci}) the characters specify the
\w-families only up to conjugation.
 \footnote{~Moreover, further degeneracies are absent at least
if \w\ is maximal.}
 In more physical terms, both $\sx^2$ and $(\sx\tX)^3$ correspond to a
combined `space' and `time' reflection (since, while they leave the
homology cycles of the torus invariant, they invert their orientation)
and hence should be equivalent to a charge conjugation.

Having arrived at the same structure in two different settings, it is
natural to speculate that, as already anticipated in the notations, the
matrix $S=\dy$ of equation (\ref{dm}) coincides with the modular matrix
$S$ that implements the map \sx\ on the characters.  This is indeed true, so
that for instance the relation (\ref{df}) translates to
the {\em Verlinde formula\/} \cite{verl2}
  \be  \nijk=\sumi l\frac{S_{il}S_{jl}(S^{-1})_{kl}}{S_{0l}} \,.  \labl{vc}
In particular, the \fra\ of any rational \cft\ is a modular \fra.

The formula (\ref{vc}) has been proven
 \footnote{~It should be stressed that to arrive at this result,
the characters must be those with respect
to the maximally extended chiral algebra; otherwise, as mentioned in
the previous section, the \furu s would not be unambigously defined.}
by performing certain formal
manipulations of two-point correlators on the torus \ttau\ that
is described in terms of $\tau$ as a parallelogram, with opposite edges
identified, whose corners are at 0, 1, $\tau$ and $\tau+1$.  These
manipulations lead \cite{verl2} to a relation like (\ref{vc}) with
certain integers \nijk.  Further, it can be shown \cite{mose3,diVe} that (as a
consequence of the pentagon identity applied to three-point correlators
on the torus) these integers are indeed the \frc s of the theory.  An
alternative proof of the formula is based on the connection
between three-dimensional topological
field theory, ribbon Hopf \alg s \cite{retu}, and \twodim\ \cft\
(in the topological setting, the matrix
elements $S_{ij}$ are obtained as expectation values of Wilson lines)
\cite{tura,witt27,frki,frga,alco2,crho}.

By the proof of the Verlinde formula
(\ref{vc}), the identification of the diagonalization
matrix $Y$ with the modular matrix $S$ follows from comparison with
(\ref{df}), provided that the (symmetric, unitary, etc.)
diagonalization matrix is unique.
As remarked at the end of section 2, this condition is almost always
fulfilled; in fact, no modular \fra\ is known for which the
diagonalization matrix cannot be fixed completely.

Let me now come to the arguments in favor of the assertion that
the transformation \sx\
acts on the characters by matrix multiplication.  One can think of the
characters $\chi$ as the chiral blocks for the zero-point function
$Z= \langle\bfe\rangle_\tau$ on \ttau, the so-called partition function
of the theory.  If the \cft\ is regarded as the vacuum configuration of a
relativistic string, then the partition function is closely related to
the path integral for the vacuum-to-vacuum transition amplitude of the
string theory. In
order to be able to properly fix the gauge symmetries of this amplitude,
the partition function must be modular invariant (see e.g.\
\cite{frms2,LUth,Sche}).
 \footnote{~To be precise, modular invariance is a sufficient condition.
It is necessary only insofar as no consistent treatment of the gauge
degrees of freedom of a non-modular invariant theory has ever been
conceived.}
 By similar arguments \cite{card3}, modular invariance is a natural
property of a
\twodim\ statistical mechanics system at the critical point of a second
order phase transition.  Together with the associativity of the \opa, the
requirement of modular invariance of $Z$ implies \cite{frsh} that the
characters span a module over the modular group.  Based on the experience
with these important manifestations of \cft, it has become common to
require modular invariance more or less axiomatically
\cite{nahm-2,frqs2} (and apply this constraint e.g.\
\cite{caiz2,scya6} to the classification of \cfts). More recently,
a general
argument for modular invariance in \cft\ has been given in
\cite{nahm7}; it employs the C$^*$-\alg ic approach to \twodim\ field
theory, thereby relating in particular the partition function to a thermal
state (at complex inverse temperature $\beta$ equal to $-2\pi\ii$ times
the modular parameter $\tau$, as is implied by interpreting $\exp(2\pi\ii
\tau L_0)$ as $\exp(-\beta H)$ \cite{Fust}) over the
field algebra. The argument involves nontrivial results about
the structure of the field algebra, which is taken to be a
von Neumann algebra of type I
 \footnote{~In this context one should note that, as is proven in
\cite{gafr}, under rather
general assumptions the observable algebra of a \cft\ is isomorphic
to the hyperfinite von Neumann factor of type III$_1$.}
 (such as the statement that any thermal state is a linear combination
of the thermal states that are associated to the irreducible \rep s of
the observable algebra, which are \cite{frrs2,gafr,jf20} in one to one
correspondence with the primary fields).

At this point it is worth recalling that most of the structural
elements of \twodim\ \cft\ are not particular to conformally invariant
theories, but are rather generic properties of \twodim\ \qft, as is
especially transparent in the C$^*$-algebraic approach to local quantum
physics
\cite{frrs}. Somewhat surprisingly, even the  relation with the modular
group is already present in this much more general context.
Namely, in the algebraic framework the
\furu s describe the composition of superselection sectors, which
can be expressed in terms of the composition of certain endomorphisms
of the local algebras of observables. It turns out
\cite{rehr2,rehr4,frga} that the matrix elements of the matrices
$Y$ and $T$ introduced in section 2 can be entirely described
 \footnote{~In particular, $Y$ is the so-called monodromy matrix.
For its definition one needs the concepts \cite{frrs} of
the `statistics operator' and the `left inverse' associated to
an endomorphism.}
 in terms of these endomorphisms, and from general properties
of the endomorphisms it follows
 \footnote{~Here a regularity property has to be assumed, which in
the conformal case corresponds to the assumption that the symmetry
algebra is maximal.}
 that the modularity constraints (M1) and (M2),
and hence the defining properties (\ref I) and
(\ref J) of the modular group, are satisfied. Moreover, for a
{\em conformal\/} theory, $T$ indeed implements the modular
transformation $\cal T$, by definition of the characters. What
is less clear is whether one can prove in a purely C$^*$-algebraic
setting that the matrix $Y$ implements the modular transformation
$\cal S$, and whether $Y$ and $T$ possess any geometric interpretation
(similar to, say, the modular operators of Tomita\hy Take\-saki
theory \cite{schr4,jf20}) for non-conformal theories as well.

According to the restriction (\ref H) on fusion dimensions, the
smallest possible fusion dimension is 1.
Generators $\phi_i$ with fusion dimension equal to one are
therefore particularly interesting; they are called
{\em simple currents.} It follows from the fact that
the quantities $\ell_i^{(n)}:=S_{in}/S_{0n}$ furnish one-dimensional
\rep s of the \fra, i.e.\ obey the sum rules
$\ell_i^{(n)}\ell_j^{(n)}=\sumi k\nijk\,\ell_k^{(n)},$ that the fusion
product of a simple current $\phi_i$ with any primary field $\phi_j$
contains only a single primary field. Accordingly
one may write $\phi_i\star\phi_j=\phi_{i\star j}$.  It can be shown
\cite{scya,intr} that the $S$-matrix elements corresponding
to fields that are related by the fusion with a simple current are
equal up to a root of unity, $S_{jk}=\exp(2\pi\ii q_{ij}/N_i)\,S_{j,
i\star k}$, with $q_{ij}$ an integer determined by the \cdim s of $\phi_i$,
$\phi_j$, and $\phi_{i\star j}$, and $N_i$ the order of the simple
current, i.e.\ the smallest positive integer such that a multiple fusion
product of $N$ factors of $\phi_i$ produces the unit $\phi_0$.
As a consequence, simple currents play an important role in the
construction of modular invariant partition functions, as well as for
the so-called field identifications which arise in the coset
construction of \cfts\ \cite{scya,scya6,gasc2}.

\subsect{\wzwts}

An important class of rational \cfts\ are the so-called (unitary)
\WZW\ (WZW) theories \cite{knza,gewi}. All known rational \cfts~are
closely related to appropriate \wzwts, namely via the coset construction
\cite{goko} or via Drin\-feld\hy So\-ko\-lov Hamiltonian reduction
\cite{drso,fortw3}. For \wzwts\ the symmetry algebra \w\ (and
$\overline{\cal W}$ as  well) is the semidirect sum of the Virasoro algebra
with an untwisted affine Lie algebra \ghat. The latter is the Lie algebra
with generators $\{K\} \cup \{\jam\mid m\in\zet,\ a=1,2,...\,,{\rm dim}
\,(\g)\}$ and brackets
  \be  [\jam,J^b_n]=\fabc\,J^c_{m+n}+\kappa^{ab}\,\delta_{m+n,0}\,K,
  \qquad  [\jam,K]=0; \ee
here $\kappa^{ab}$
and \fabc\ denote the Killing form and structure constants of
a semisimple Lie algebra $\g\subset\ghat$, namely the subalgebra
generated by the zero modes $J^a_0$, i.e.\ the horizontal subalgebra of \ghat.

In addition, for a \wzwt\ the Virasoro generators
$L_n$ are quadratic expressions in the
\ghat-generators \jam. Also, all \hwm s of \ghat\ which appear
in a given theory possess the same eigenvalues $k$ and $c$ of the central
generators $K$ and $C$, and these numbers
are related by $c=k\,{\rm dim}(\g)/(k+h)$,
with $h$ the dual Coxeter number of \g. $k$ is called the
level of \ghat; for unitarity it must be a positive integer
(I normalize the inner product on the weight space of \g\
such that the highest root \tta\ of \g\ satisfies $(\tta,\tta)=2$).
The spectrum $\{\phi_{\Lambda}\}$ of primary fields can be
labelled by \hw s $\Lambda$ of finite-dimensional \g-modules
whose inner product with the highest root is not larger than the
level, i.e.\
  \be  I=I_k\equiv I(\g,k)= \{\Lambda\in(\zetpluso)^{{\rm rank}
  (\g)}_{}\mid (\Lambda,\tta)\leq k\}\,,  \labl{ik}
and the conjugation of primary fields corresponds to the
conjugation of highest \g-weights, $(\phi_\Lambda)^+_{}=\phi_{\Lambda^+_{}}
^{}.$

For \wzwts\ the modular matrix $T$ takes the form
  \be  T^{}_{\Lambda\Lambda'}=\delta^{}_{\Lambda\Lambda'}
  \,\exp\left[\pi\ii\left((\Lambda,\Lambda+2\rho)-k\,{\rm dim}(\g)/12
  \right)/(k+h)\right],   \ee
with $\rho$ the Weyl vector of \g. Also, as a consequence of the fact that
the characters \chil\ are just specializations of the ordinary affine
characters of \ghat-modules, the matrix $S$ is given by the
following {\em Kac\hy Pe\-ter\-son formula\/} \cite{kape3}:
  \be  S^{}_{\Lambda\Lambda'}={\rm const} \cdot \prod_{\alpha>0}
  \sin\mbox{\LARGE(}\frac{\pi(\Lambda+\rho,\alpha)}{k+h}\mbox{\LARGE)}
  \cdot\frac{ {\displaystyle\sum_{w\in W}} \sigma(w)\,
  \exp\mbox{\Large[}-\frac{2\pi\ii}{k+h}(\Lambda+\rho,w(\Lambda'+\rho))
  \mbox{\Large]} } { {\displaystyle\sum_{w\in W}} \sigma(w)\,
  \exp\mbox{\Large[} -\frac{2\pi\ii}{k+h}
  (\Lambda+\rho,w(\rho))\mbox{\Large]} } \,;  \label{kp} \ee
here the product is over the positive roots of \g, the sums are
over the Weyl group $W$ of \g, and $\sigma(w)$ stands for the sign of
the Weyl group element $w$.

In the notation appropriate to \wzwts, the \furu s are written as\,
$\phi_\Lambda\star\phi_\Lambda'=\sumik{\Lambda''}{\cal N}_{\Lambda
\Lambda'}^{\;\;\Lambda''}\phi_{\Lambda''},$ and the Verlinde formula reads
  \be  {\cal N}_{\Lambda \Lambda'}^{\;\;
  \Lambda''}=\sumik\mu\frac{S_{\Lambda\mu}S_{\Lambda'\mu}
  S^*_{\Lambda''\mu}}{S_{0\mu}}.    \labl{vW}
Unfortunately,
due to the summation over the Weyl group the calculation of
the Kac\hy Pe\-ter\-son $S$-matrix is cumbersome for `large' \alg s,
so that the Verlinde formula is in itself of limited use.
But fortunately there also exist other possibilities
of computing the \wzw~\furu s; all of them are in close relation
with the \rep\ theory of semisimple Lie \alg s,
and in particular require the knowledge of the tensor product
multiplicities of \g, which (deviating, for further convenience,
slightly from the notation that was used in (\ref a)) will be denoted by
$\nlllb$.

One of these algorithms is the so-called depth rule
\cite{gewi,kmsw} which seems however also quite involved
already for modestly large \alg s, and which for general level
so far has only been applied to $\g=A_1$ \cite{gewi} and to $\g=A_2$
\cite{cumm,bemw2}.  Another one expresses
\cite{KAc3,walt3} the \wzw~\frc s $\nlll\equiv\nlll(k)$
as a weighted sum over the tensor product multiplicities
$\overline{\cal N}_{\Lambda \Lambda'}^{\;\;\;\,\mu}$,
  \be  \nlll =\sum_{\hat w\in\hat W}\sigma (w)\,\overline{\cal N}
  _{\!\Lambda \Lambda'}^{\;\; \hat w(\Lambda'')}.  \labl{kw}
Here $\hat W$ denotes the horizontal projection
of the Weyl group of \ghat, i.e.\ $\hat w\in\hat W$ corresponds to
a pair $(w,\beta)$ with $w\in W$
and $\beta$ an element of the coroot lattice of \g, acting as
$\hat w(\Lambda) =w(\Lambda+\rho)-\rho+(k+h)\beta.$
While $\hat W$ is an infinite group, for any triple $(\Lambda,\Lambda',
\Lambda'')\in(I_k)^3_{}$ the sum in (\ref{kw}) contains only a finite
number of
non-vanishing terms, and there exists a finite algorithm to evaluate
the formula. As a consequence, the formula
is easily implemented on a computer. If \g\
is a classical Lie algebra, it can also be translated
into a Young-diagrammatic prescription, as has been
done in \cite{cumm} for $\g=A_r$ and $\g=C_r$.

The result (\ref{kw}) is a rather straightforward
consequence of the Kac\hy Pe\-ter\-son formula for $S$; nevertheless it is
instructive to mention a few details \cite{walt4,fugp2,fuva2} of its
proof. Namely, on one hand the numbers
$\ell_{\Lambda}^{\;(\mu)}:=S_{\Lambda\mu}/S_{0\mu}$ obey the
\rep\ property
  \be  \ell_{\Lambda}^{\;(\mu)} \ell_{\Lambda'}^{\;(\mu)}
  =\sumik\lapp{\cal N}_{\Lambda \Lambda'}^
  {\;\;\;\Lambda''} \ell_{\Lambda''}^{\;(\mu)} .  \labl N
On the other hand, the Kac\hy Pe\-ter\-son formula implies
$\ell_{\Lambda}^{\;(\mu)}=\chilb(2\pi\ii(\mu+\rho)/(k+h)),$
with \chilb\ the character of the \g-module $\L_\Lambda$ (not to be
confused with the specialized affine character \chil), and hence
the character sum rule for tensor products of \g-modules implies
  \be  \ell_{\Lambda}^{\;(\mu)} \ell_{\Lambda'}^{\;(\mu)}
  =\sum_{\Lambda''\in I_\infty} \overline{\cal N}_{\Lambda \Lambda'}^
  {\;\;\;\Lambda''} \ell_{\Lambda''}^{\;(\mu)} .  \labl L
The two sum rules are compatible with each other because
$\ell_{\lambda}^{\,(\mu)}=\sigma(w)\,\ell_{\hat w(\lambda)}^{\,(\mu)}$
for any $\hat w\in\hat W$, and because for given
$\Lambda$ there is at most one $\hat w\in\hat W$ such that
$(\hat w(\Lambda),\tta)\leq k$. Finally, there are no further
relations among the $\ell_{\Lambda}^{\,(\mu)}$, i.e.\
the set $\{\ell_{\Lambda}^{\,(\mu)}\mid\Lambda\in I_k\}$ is
linearly independent (otherwise the columns of $S$
were dependent, in contradiction to $S^4=\one$). Putting these results
together, the formula (\ref{kw}) follows.

The prescription (\ref{kw}) is particularly simple if
$(\la,\tta)=1$. In this case it follows that
$(\lapp,\tta)\leq k+1$ for all $\lap,\lapp\in I_k$ with
$\nlllb\neq0$, and
  \be   \nlll=\left\{\begin{array}{lll} \nlllb &{\rm for}&
  (\lapp,\tta)\leq k, \\[.9 mm] 0 &{\rm for}&
  (\lapp,\tta)=k+1. \end{array} \right. \labl{k1}
In other words, the \furu s of fields satisfying the constraint
$(\la,\tta)=1$
are obtained from the corresponding tensor products of \g-\rep s
by merely removing the couplings to fields with
$(\lapp,\tta)=k+1$.

{}From the results of the algorithms for \wzw\ \furu s, one
can conversely deduce the matrix $S$, in the manner
explained at the end of the section 2. Actually \cite{fuva3}, as
input for the diagonalization procedure one usually needs only a
single (non-trivial) fusion matrix, and hence only a
single matrix of tensor product coefficients. Since the
computation of the latter is the major obstacle that
limits the use of the depth rule or of (\ref{kw}), the
most efficient way to proceed is to use first the depth rule or
(\ref{kw}) for the computation of a single fusion matrix,
then diagonalize, thereby obtaining the modular matrix $S$,
and then calculate the remaining fusion matrices via the Verlinde
formula.

As can be seen by investigating the depth rule \cite{kmsw},
to each of the \nlllb\ distinct couplings $(\Lambda,\Lambda',
\Lambda'';p)$ that arise at some level of a \wzwt, one can associate
a `treshold
level' $k_p\equiv k_{\Lambda,\Lambda',\Lambda'';p}$ such that
the coupling is present for all levels $k\geq k_p,$ but absent for
all $k<k_p.$ In particular, $p$ takes the values $p=1,2,...\,,\nlllb$,
the \nlll\ are majorized by the tensor product coefficients,
  \be  0\leq \nlll(k) \leq \nlllb   \labl{ll}
(this it is not manifest in the formula (\ref{kw})), and
  \be  \begin{array}{lll} \nlll(k)=0 &{\rm for}& k<\min \{k_p\mid
  p=1,...\,, \nlllb\}\,,   \\[2 mm] \nlll(k)= \nlllb
  &{\rm for}& k\geq\max\{k_p\mid p=1,...\,, \nlllb \}  \,. \end{array}\ee
As an illustration, I list in the following table
three specific \frc s of the $E_8$ \wzwt\ at levels $\geq4$ (for
smaller levels, the relevant field $\phi_\lles$ does not belong to the
spectrum):
  \be  \begin{tabular}{l|cccl} \multicolumn{1}{c|}{$k$}
   &4&5&${\scriptstyle\geq}6$\\[1.2 mm]\cline{1-4}
  &&\\[-2.7 mm] ${\cal N}_{\lles,\lles,\lle}$ & 1 & 2 & 2 \\[1.1 mm]
  ${\cal N}_{\lles,\lles,\lls}$ & 2 & 3 & 3 \\[1.1 mm]
  ${\cal N}_{\lles,\lles,\lles}$& 2 & 8 & 9 &. \end{tabular}  \ee

Another property of \wzw\ \furu s that can easily be verified is
\cite{jf10,fugp2,gepn9} that
  \be   {\cal N}_{\omega(\Lambda)\,\omega'(\Lambda')}^{\ \ \ \ \ \ \omega
  \omega'(\Lambda'')}=\nlll  \ee
where $\omega,\,\omega'$ is any pair of automorphisms of the extended
Dynkin diagram of \g\ that are not automorphisms of the unextended
Dynkin diagram.

Let me finally mention that
if the level is taken to be a non-real complex number or a
real number smaller than $-h$ (which means in particular that
the theory is non-unitary), then \cite{kalu} it is possible
to define a tensor product of \ghat-modules
 \footnote{~The \ghat-modules must obey the usual restriction
that their subspaces at any fixed grade are \findim.}
 that does not change the level and hence is in this respect similar
to the \cft\ \furu s.  Further, there does exist a compatible tensor
functor from the monoidal category with
this tensor product to the category of \findim\ \g-modules, and
hence also such a functor to the category of \findim\ vector spaces.

\subsect{Polynomial rings and fusion potentials}
\subsection{Fusion rings and polynomial rings}

Any \findim\ commutative associative ring, and hence in particular any
rational fusion ring \ca, can be presented as the quotient of a
free polynomial ring by some ideal. Namely, consider the elements
$\phi_i$ of a canonocal basis of \ca\ as formal variables; then one has
  \be  \ca \cong \zet[\phi]/{\cal J} \,, \labl{cZ}
with $\zet[\phi]\equiv\zet[\phi_1,\phi_2,...\,,\phi_\II]$ the ring of
polynomials (with integral coefficients) in these variables,
and $\cal J$ the subring generated by the fusion
relations, i.e.\ by the polynomials
  \be  P_{ij}(\phi):=\phi_i\phi_j-(\phi_i\star\phi_j)(\phi)=
  \phi_i\phi_j-\sumi k\nijk\phi_k \,.\ee
Due to commutativity and associativity, $\cal J$ is
a two-sided ideal of $\zet[\phi]$, and as a consequence
dividing it out as in (\ref{cZ}) is a well-defined procedure
so that indeed (\ref{cZ}) can be used as a definition of the
fusion ring.

Consider now the $\phi_i$ as complex variables rather than as
formal indeterminates.  Then owing to the \rep\ property (\ref E) of
\furu\ eigenvalues, one has
  \be  P_{ij}(\phi)=0  \qquad {\rm iff}\ \  \phi_l=\nv kl
  =\yy kl/\yy k0 \ \ {\rm for\ all}\ l\in I
  \mbox{ and some } k\in I\, . \labl{63}
As a consequence, $\cal J$ may also be characterized as being
generated by \Ii\ polynomials \pk\phi\ with the property that
$\pk\phi=0$ iff $\phi_i=\yy ki/\yy k0$ for all $i\in I$.
These polynomials can be chosen as $\pk\phi=\proi j(\phi_j-\nv kj)$,
but because not all of the $\II^2$ numbers \nv li are distinct,
one may also restrict the product to the terms with certain appropriate
eigenvalues \nv kj. In particular, for $k=0$ one can replace
${\cal P}_0(\phi)=\proi j(\phi_j-\cd j)$ by
${\cal P}_0(\phi)=\phi_0-1$, thereby effectively setting $\phi_0$
equal to one and hence eliminating it as an independent variable.
Often the quotienting by $\cal J$ effectively eliminates some further
generators $\phi_i$ as well; for example, as it turns out
all primary fields of the fusion ring of the
$A_1$ \wzwt\ are generated by the single element $\phi_{\la_{(1)}}$.
It is then natural to consider \ca\ as obtained by quotienting
the free polynomial ring in the independent variables $\phi_{j_i}
=:x_i$, $i=1,2,...\,,n$.  Thus the fusion ring is written as
  \be  \ca\cong\zet[x]/{\cal J} \,,  \labl{zxi}
where now $\cal J$ is the ideal generated by some polynomial constraints
$\pk x =0$, $k=1,2,...\,,n$,
in the variables $x_i$, and these polynomials $\pk x$,
vanish simultaneously iff for some $l\in\{1,2,...\,,n\}$
one has $x_j=\mv lj$ for all $j=1,2,...\,,n$, where \mv lj
are the eigenvalues of $\pireg(x_j)$.

\subsection{Local rings}

Particularly interesting is certainly the situation where
the polynomials \pk x are integrable, which means that they
are derived from some potential $V(x)$ in the sense that
  \be  \pk x =\frac\partial{\partial x_k}\,V(x) \ee
for all $k$. In other words,
  \be  \ca\cong\zet[x]/{\rm d}V(x) , \labl v
i.e.\ \ca\ is the {\em local ring\/} of the potential $V$.

Given any fusion ring \ca\ one may attempt to construct a presentation
as a local ring as follows. Consider for the moment \ca\ as an algebra
over the field \complex\ (actually, an appropriate finite algebraic
extension of {\dl Q} already does the job), and take an element $\tilde x
\in\ca\,$ for which all eigenvalues of $\pireg(\tilde x)$ are distinct;
as pointed out at the end of section 2, such an element does exist.
Now instead of the canonical basis $\{\phi_i\}$ of \ca, consider
a particular basis $\cal B$ that contains $\tilde x$, namely ${\cal B}=
\{\phi_0, \tilde x,\phi_2,\phi_3,\ldots,\phi_\II\}$; here it is assumed
that the coefficient $a_1$ in the decomposition
  \be  \tilde x=\sum_{j=0}^{\II-1}a_j\,\phi_j  \labl{xaf}
of $\tilde x$ with respect to the canonical basis does not vanish, which
can always be accomplished by appropriate labelling of the
generators $\phi_i$. Next eliminate $\phi_0$ and possibly
further elements of $\cal B$ (but {\em not\/} $\tilde x$, even if this
were possible) in the manner described above. Writing from now on
\tx\ for $\tilde x$, one arrives at a presentation of the form
(\ref{zxi}) of \ca\ in terms of the variables $\{\tx,x_2,x_3,\ldots,
x_n\}$ (the tilde on the first variable is kept in order to emphasize
that, generically, in contrast to the other variables it is not
an element of the canonical basis).

Now denote the eigenvalues of $\pireg(\tx)$ by \mV k, and, as before,
those of $\pireg(x_i)$ by \mv ki\ for $i=2,3,...\,,n$.
Then make the ansatz \cite{ahar}
  \be  \begin{array}{ll}  V(\tx,x_2,...\,,x_n)= & {\displaystyle
  -\sumI i\mv i2\int^{\tilde x_1}
  \!\!{\rm d}t\, \proine ji (t-\mV j) + x_2\,\proI i \mvX i} \\[1.7 mm]
  & \displaystyle{ + \onehalf \sumI i \sum_{j=3}^n \mvx jij ^2_{}
  \proine li \mvX l } \,. \end{array} \labl{ah}
The partial derivatives of (\ref{ah}) read
  \be   \begin{array}{l}  \displaystyle{\dtx V(x)=\sumI i \mbox{\large (}\!
  \proine ji \mvX j \mbox{\large )} \cdot \mbox{\large [} \mvx2i2 + \onehalf
  \sum_{j=3}^n \mvx jij ^2_{} \sumine li \mvX l ^{-1}_{}
  \mbox{\large ]}\, ,} \\ {}\\[-2.2 mm] \displaystyle{
  \dx2 V(x) = \proI i \mvX i\,, } \\ {}\\[-2.2 mm] \displaystyle{
  \dx j V(x) = \sumI i \mvx jij \proine li \mvX l \quad {\rm for}
  \ \ j\geq3.} \end{array} \ee
Thus requiring $\dX2x=0$ enforces $\tx=\mV k$ for some $k\in I$;
as all $\mV l$ are distinct, one has $\dX j{\tx=\mV k;x_2,...\,,x_n}
=\mvx jkj\proinE lk\mvX l$ for $j\geq3$, so that $\dX jx=0$ enforces
$x_j=\mv kj$; finally, $\dtX{\tx=\mV k;x_2;x_3=\mv k3,...\,,x_n=\mv kn}
=\mvx 2k2\proinE lk\mvX l$, so that $\dtX x=0$ enforces $x_2=\mv k2$.

In short, one finds that $\dX ix=0$ iff $x_j=\mv jk$ for all
$j\in I$ and some $k\in I$, i.e.\ the standard property of the
constraints \pk x described above.
Moreover, clearly the \Ii\ polynomials $1$, $\tx$, $(\tx)^2$, \ldots,
$(\tx)^{\II-1}$ are linearly independent over \complex\ (whereas $(\tx)^
\II$ can be expressed through these owing to the explicit form of the
constraint $\dX2x=0$).
Since \Ii\ is the dimension of the fusion ring, these polynomials
provide a basis, and hence a reconstruction of the ring from
the potential is possible. In particular it follows that there
exist polynomials $Q_j(\tx)$ and $R_{ij}(\tx,x_2,...\,,x_n)$
for $j=2,3,...\,,n$ and $i=1,2,...\,,n$ such that
  \be  x_j=Q_j(\tx)+R_{1j}(x)\,\dtx V(x)+\sum_{i=2}^n R_{ij}(x)\,\dx j
  V(x), \labl{qr}
i.e.\ such that $x_j=Q_j(\tx)$\,mod\,d$V$.
As a consequence, as an algebra over \complex, \ca\ can indeed be
presented as the {\em local algebra\/} $\complex[\tx,x_2,...\,,x_n]
/{\rm d}V$. Whether as a ring \ca\ can be written analogously as
$\zet[\tx,x_2,...\,,x_n]/{\rm d}V$ is, however, difficult to decide, as
one would have to show that all coefficients in the polynomial $V$ as
defined in (\ref{ah}) are rational numbers
 \footnote{~There are however arguments (see \cite{ahar}, and O.\
Aharony, private communication) which show that
the coefficients can indeed be taken as rational.}
 (and hence, with appropriate over-all normalization of $V$,
integers); if they are not, then \ca\ is not a well-defined local ring
over \zet.

Moreover, by assumption there exist polynomials $p_i$ such
that $\phi_i=p_i(\tx,x_2,...\,,x_n)$ for  $i=2,3,...\,,\Ii$, and
hence (\ref{xaf}) and (\ref{qr}) imply
  \be  \begin{array}{ll}\phi_1\! &=\;a_1^{-1}\,(\tx-\sumine i1 a_i\,\phi_i)
  \\ & =\; a_1^{-1}\,(\tx-a_0-\sum_{i=2}^\II a_i\,
  p_i(\tx,Q_2(\tx),...\,,Q_n(\tx)) \;{\rm mod\,d}V
  \\{}\\[-1.4 mm] & =:Q_1(\tx)\;{\rm mod\,d}V  \end{array} \ee
for the generator $\phi_1$ of the canonical basis.
Note that the coefficients $a_i$ introduced in the decomposition
(\ref{xaf}) are generically not rational. Thus even if \ca\ could
be presented as a local ring in the variables
$\tx,x_2,...\,,x_n$, it would in general still not be a local ring
in the canonical variables $x_1\equiv\phi_1,x_2,...\,,x_n$.

In short, for any fusion algebra
of a rational \cft\ there exists a presentation of the form
  \be  \ca\cong\complex[x]/{\rm d}V(x)  \ee
analogous to (\ref v). It should, however, be realized that the
potential $V$ appearing here is typically far from being unique.
Namely, first, for a generic \fra\ the space of elements $\tilde x$
for which all eigenvalues of $\pireg(\tilde x)$ are distinct is of
dimension larger than one. Second, for a fixed choice of $\tilde x$
with this property, typically several coefficients $a_j$ in the
expansion (\ref{xaf}) are non-vanishing, so that instead of
trading $\tilde x$ for $\phi_1$, one could trade it for other
elements of the canonical basis as well.
With respect to both of these ambiguities,
different choices will usually lead to different presentations
of the fusion ring. As a consequence, $V$ must merely be considered as a
condensed description of the fusion ring, and usually should not be
expected to possess any independent meaning (such as, for instance, as
the Landau\hy Ginzburg potential of some lagrangian field theory).
But perhaps the situation is special if the presentation
as a local algebra is in terms of canonical generators of the
fusion ring (i.e.\ if $\tx=x_1$ in the computations above).
The latter situation is realized for the $A_r$ \cite{gepn9}
and the $C_r$ \cite{bmrs,gesc} \wzwts\ (see section 7 for
some details on the $A_r$ fusion rings), and -- as has been advocated
\cite{cres}, based on the relation \cite{witt27} between the canonically
quantized three-dimensional Chern\hy Simons gauge theory and \wzwm s
-- persists for all other \wzwts\ as well.
Also \cite{ahar}, for the $c<1$ unitary minimal models, there
exists a presentation of the fusion ring as a local ring in
two canonical generators (namely, $x_1=\phi_{(1,2)}$ and
$x_2=\phi_{(2,1)}$ in the notation of \cite{bepz}).

To conclude this subsection, let me point out that, although
for a local ring the complete information on the ring structure
is encoded in the single function $V(x)$,
this information is generically not sufficient to reconstruct the
fusion ring in its canonical basis $\{\phi_i\}$.

\subsection{One-variable fusion potentials}

A special class of local rings are those derived from the free
polynomial ring in a single variable $x$, for which the
integrability condition is trivial. The present subsection is
devoted to this particular situation. Thus assume that there exists
an element $\phi$ of a canonical basis together with \Ii\
polynomials $p_j(\phi)$ that are linearly independent over
\complex, such that
  \be  \phi_i=p_i(\phi)  \labl{84}
for all $i\in I$. In particular, $p_0(\phi)=1$, and, numbering the
generators such that $\phi_1=\phi$, the first polynomial is
$p_1(\phi)=\phi$. By evaluation of (\ref{84}) in the regular \rep,
one has \cite{capr2}
  \be  \cn i=p_i(\cn1),  \labl{85}
which implies that the \furu\ eigenvalues obey
  \be  \nv ji =p_i(\nv j1) . \labl{86}
Moreover, because of the vanishing condition (\ref{63}) the ideal
$\cal J$ in (\ref{zxi}) is generated by
  \be  {\cal P}(\phi):=\prod_j(\phi-\nv j1) .\ee
Without changing the ring structure, the product on the right hand side may
be restricted to those $j\in I$ that correspond to {\em distinct\/}
eigenvalues \nv j1\ of \cn1. But if the number of these were smaller
than \Ii, then the assumption that in $\zet[x]/{\cal J}$ there exist
\Ii\ independent polynomials $p_i$ could not be fulfilled. One concludes
\cite{dizu2} that all eigenvalues of
\cn1\ are distinct, i.e.\ non-degenerate (as a consequence,
the information contained in \cn1\ is sufficient to  fix uniquely
the eigenvectors, and hence the diagonalization matrix \dm\ via
(\ref{23})).

In the above, it was assumed that the coefficients of the
polynomials $p_i$ take values in \zet. Note, however, that
it follows already from (\ref{85}) (along with the fact that
$\nijk\in\zet$) that the coefficients of the $p_i$ are
rational. One can show that the fact that the eigenvalues of \cn1\
are non-degenerate is not only necessary, but also sufficient
for (\ref{84}) to hold for independent polynomials with rational
coefficients. Namely \cite{dizu2}, define for any $j\in I$ the
polynomial $p_j$ as the unique polynomial of order at most $\II-1$
that obeys (\ref{86}) (and hence (\ref{85})). As a function of
one complex variable $\phi$, $p_i$ then satisfies (\ref{84}) at
the points $\phi=\nv j1$. If all the \Ii\ eigenvalues \nv j1 are distinct,
this implies that (\ref{84}) is true for any value of $\phi$.

There exist \cfts\ for which
all of the fusion matrices possess degenerate eigenvalues, e.g.\
the unitary minimal models (except the Ising model).
But even for such theories it is still possible that the relation
(\ref{84}) is fulfilled, albeit with coefficients in
an algebraic extension of {\dl Q} by some of the fusion eigenvalues
(this happens for instance for all the minimal conformal models
\cite{dizu2}).
As a consequence, while not being independent over \complex, the
polynomials $p_i$ are still independent over {\dl Q}. This makes such
polynomial presentations again interesting, as independence over
{\dl Q} is all that is needed for the presentation to be faithful,
and hence for a reconstruction of the fusion rules
from the potential $V$ and from the polynomials $p_i$.

To see how such polynomial presentations arise,
fix some $x\in\ca$, and denote by $\mu_j$, $j=0,1,...\,,\mm$
($\mm\leq\Ii-1$), the distinct eigenvalues of \pix. Since \pix\ is
linear combination of the matrices \cn j, its eigenvectors $v_i$
coincide with those appearing in (\ref{23}). Having interpreted
the generators of \ca\ as complex variables, the eigenvectors $v_i$
may be formally considered as elements of \ca. Doing so, one finds that
  \be   v_i=\sumi j \dy^*_{\!ij}\,\phi_j  , \labl{a0}
since by use of the \furu s this correctly reproduces the eigenvalue
equation in the form $\phi_l\star v_j=(\yy jl/\yy j0)\,v_j$.
According to (\ref{a0}), the eigenvectors $v_i$ correspond up
to normalization to the minimal idempotents (\ref{ef}) of
\ca. In particular,
  \be  v_i\star v_j=\delta_{ij}v_i/\yy i0.  \labl{a2}

Assume now that there exists  a polynomial presentation of \ca, as an
algebra over \complex, in terms of the element $x$. Then
\cite{ahar} the constraint to be imposed on the free algebra $\complex[x]$
must be the minimal polynomial of \pix, i.e.
  \be  V'(x)=\prom j (x-\mu_j) .  \labl{a1}
By assumption, in particular each of the elements (\ref{a0})
must be representable as a polynomial in $x$. Ordering the
eigenvectors in such a manner that $v_i$ has eigenvalue $\mu_i$
for $i=0,1,...\,,\mm$, the eigenvalue equations of these
elements read $(\pix-\mu_i\one)\cdot v_i(\pix)=0$, which by comparison
with (\ref{a1}) implies
  \be  v_i(x)\propto \promne li \frac{x-\mu_l}{\mu_i-\mu_l }.
  \labl{a3}
Inserting this result into (\ref{a2}), it follows by
specializing to $i=j$ and $x=\mu_i$ that the constant of
proportionality must be either zero or equal to $1/\yy i0$. Moreover,
using (\ref{a2}) with $i\neq j$ along with $\sumi j \yy0j v_j(x)=
\phi_0(x)=1$, one then deduces that in fact
  \be  v_i(x)=(\yy i0)^{-1}\promne li \frac{x-\mu_l}{\mu_i-\mu_l }
  \qquad {\rm for} \ \ i=0,1,...\,,\mm , \labl{a4}
as well as
  \be \mbox{\hspace{3.4 em}} v_i(x)\equiv 0 \mbox{\hspace{7 em}}
  {\rm for} \ \ i=\mm+1,\mm+2,...\,,\Ii-1. \ee
Thus the polynomial \rep\ of the generators reads \cite{ahar}
  \be  \phi_i(x)=\sumi j \yy ij\,v_j(x)= \summ j \frac{\yy ij}
  {\yy 0j} \promne lj \frac{x-\mu_l}{\mu_j-\mu_l} \,. \ee
It remains to be checked whether these polynomials are independent
over \complex\ or at least over {\dl Q}, i.e.\ whether there
exists a linear combination
  \be  \tilde\phi(x)=\sumi i a_i\,\phi_i(x)=\sumi{i,j} a_i \yy ij\,
  v_j(x) \labl{ti}
that vanishes identically. Clearly, for arbitrary coefficients $a_i$,
  \be  \tilde\phi=\sumI i a_i \sum_{j=\mm+1}^{\II-1} \yy ij\, v_j =:
  \sum_{j=\mm+1}^{\II-1} \tilde a_j\, v_j   \ee
satisfies this constraint,
and from (\ref{a1}) and (\ref{a4}) it follows that all solutions
are of this form.
Thus, in agreement with the result above, independence over \complex\
requires that $\mm=\Ii-1$. Concerning independence over {\dl Q}, one
must investigate whether there
exist complex numbers $\tilde a_j$, $j=\mm+1,\mm+2,...\,,\Ii-1$,
for which $a_i= \sum_{j=\mm+1}^{\II-1} \dy^*_{\!ij}\tilde a_j$
is rational for any $i\in I$. So far no general answer to this question
is available. For any given fusion ring,
one may of course use the explicit form of the matrix $Y$ to analyze
the problem. A few theories for which in this manner one can show the
absence of rational solutions for the $a_i$, and hence independence
of the polynomials over {\dl Q}, are described in \cite{ahar}.

\subsection{Quasihomogeneous polynomials}

Another special class of local rings are those for which the potential $V$
is a quasihomogeneous polynomial, i.e.\ satisfies
  \be  V(\lambda^{m_1}x_1,\lambda^{m_2}x_2,...\,,\lambda^{m_n}x_n)
  = \lambda^M\,V(x_1,x_2,...\,,x_n)  \ee
with some integers $M$ and $m_1,\,m_2,...\,,m_n$ for any $\lambda\in
\complex\setminus\{0\}$. The degree of quasihomogeneity provides a quantum
number $Q$ that is additively conserved under the ring product. As a
consequence, the local ring of $V$ is either \infdim, or else is nilpotent
(in fact \cite{ARgv} the ring is \findim\ only if $V$ has an isolated
singularity at $x=0$). But the existence of nilpotent elements is not
allowed in a \fra. Namely, from the properties of the unit $\phi_0$ and the
conjugation one knows that $\phi_j\star\phi^{}_{j^+}=\phi_0+...\,,$
where the ellipsis stands for further generators, and hence
$(\phi_i\star\phi_j)\star\phi^{}_{j^+}=\phi_i\star(\phi_j^{}\star\phi^{}
_{j^+})=\phi_i+...\,,$ implying in particular that the
fusion product of any two generators cannot vanish, $\phi_i\star\phi_j
\neq0.$ To obtain the same result for arbitrary elements $x\in\ca$, it
is most convenient to write them in the basis of minimal idempotents
(\ref{ef}),
 \footnote{~That this is of course only a basis of \ca\ as an
algebra over $\scriptstyle\complex$ does not affect the argument.}
 $x=\sumi i a_i e_i$, which shows that $x^n=\sumi i(a_i)^n e_i$ for any
$n\in\zetplus$ so that $x^n\neq0$ unless $x=0$.
 \footnote{~In contrast, \fra s generically do possess zero divisors.
For example, $(\phi_0+\phi_i)\star(\phi_0-\phi_i)=0$ \,if $\phi_i$ is
a simple current of order 2.}

Thus a local algebra corresponding to a quasihomogeneous potential
does not satisfy the axioms of a \fra. But the structure is still
close enough to that of a \fra; the axioms (F1), (F2), and (F3)
hold, and it is also possible to define a conjugation $\tilde C$, but
$\tilde C$ cannot be unital.
A particularly interesting nilpotent element of a \findim\ local ring of a
quasihomogeneous potential $V$ is
  \be  \phi_c:={\rm det}_{i,j}(\partial_i\partial_jV) . \ee
$\phi_c$ is the unique element with maximal charge $Q$. Namely,
if $Q$ is normalized
such that $V$ has unit charge, i.e.\ $Q_i=m_i/M$, then the charge of
$\phi_c$ is $1-2\sum_i Q_i$; on the other hand, the polynomial $P(t)
:={\rm tr}_{\cal A}^{}(t^{MQ})$, known as the Poincar\'e polynomial of \ca,
can be shown \cite{ARgv,levw} to read $P(t)=\prod_i(1-t^{M-m_i})
(1-t^{m_i})^{-1}$, so that from the limit of large $t$ one can read
off that $1-2\sum_i Q_i$ is the largest allowed charge and appears
with multiplicity one.

A class of \cfts\ in which such structures arise are the $N=2$ super\cfts
\cite{levw}. In this context, the charge quantum number $Q$ is associated
with the $u_1$-subalgebra of the $N=2$ superconformal algebra;
as a consequence of the structure of the $N=2$ algebra, $Q$ is bounded
by the conformal dimension as $Q\leq2\Delta$ (assuming that the theory
is unitary). The elements of the nilpotent ring are then the so-called
chiral primary fields, i.e.\ those primaries for which this bound on
the charge is saturated. The
ring multiplication corresponds to the operator product in the limit of
vanishing distance, which is well-defined owing to the conservation of
the $u_1$-charge under operator products. (Moreover, it is possible to
remove all non-chiral primary fields and all descendants by an
appropriate `twisting' procedure; the dependence of the fields on
their position in \twodim\ space-time then becomes irrelevant, and the
theory obtained this way is topological, possessing only a finite number
of degrees of freedom \cite{egya2,witt27}.)
Also, the charge $1-2\sum_i Q_i$ of $\phi_c$ equals one third of the
conformal central charge. Furthermore, the conjugation $\tilde C$
amounts to forming the fusion product with the simple current $\phi_c$,
followed by conjugation in the \cft\ sense (an equivalent description
of this operation is the following: first apply the so-called
spectral flow operator to pass from the
Neveu\hy Schwarz sector to the Ramond sector of the theory,
then perform conjugation in the \cft\ sense, and afterwards flow back to the
Neveu\hy Schwarz sector). This implies in particular that
   \be  \tilde C_{i0}=\delta_{i,c}. \ee

Due to the similarities between the two types of rings, it seems natural to
expect that there exist fusion rings which are obtained from the
local ring of a quasihomogeneous potential by a certain perturbation.
It is also plausible to require that the deformation
must not change the number of critical points (counting multiplicities)
of the potential (compare \cite{vafa6}). Indeed it has been shown
for various special cases
\cite{lewa2,ceva,fein2,spie3} that the deformation of the quasihomogeneous
polynomial $V$ of the ring of chiral primary fields of an $N=2$ theory
by the polynomial corresponding to a single chiral primary
field can result in a fusion ring. Note that
if the deformation is by any integral linear combination of
chiral primaries, then \cite{intr2} the fusion ring will
contain a simple current, namely (the image under deformation of)
the element $\phi_c$.

\subsect{Example: the \slnk\ fusion ring}

As an example for \cfts\ for which some of the issues of the previous
sections can be discussed in a rather explicit manner, consider the
\wzwts\ based on one of the simple Lie algebras $\g=\sln$, at some level
$k\in\zetplus$. Denote by $I_k:=I(\sln,k)$ the index set (\ref{ik}), by
$c_\la$ the conjugacy class of $\la\in I_k$, by $\la_{(j)}$,
$j=1,2,...\,,n-1$, the fundamental weights of \sln, and by
$\la_{(0)}$ the zero weight. Then the
\furu s for any fixed choice of $n$ and of the level $k$ can be
characterized uniquely (up to isomorphism) as follows:
\begin{itemize}\ibox The generators $\phi_\la$ are indexed by
$I_k$, and \ibox the \furu s of $\phi_{\Lambda_{(j)}}$ read
 \footnote{~Because of $(\la_{(j)},\tta)=1$ for $j=1,2,...\,,n-1$,
this is a special case of (\ref{k1}).}
  \be  \phi^{}_\la\star\phi_{\Lambda_{(j)}}=\bigoplus_{\lap\in M^\la_j}
  \phi_\lap \labl{7.1}
for $j=0,1,...\,,n-1$, where
  \be  M^\la_j:=\{\lap\in I_k\mid c_\lap=c_\la+j;\; 0\leq (\lap)^i-
  \la^i\leq1\ \forall\;i=1,2,...\,,n-1\}.  \ee
\end{itemize}
That the \slnk-\furu s indeed have these properties can easily be
deduced from the formula
(\ref{kw}); alternatively, one may derive them with Young-diagrammatic
methods. That the properties are enough to specify the \furu s uniquely
can be shown by proving \cite{gona,gowe} that the abstract
fusion ring possessing these properties is unique. Incidentally,
the commutativity of the ring is not needed for the proof, but rather
can be deduced from the other properties.
(One may also verify these properties without the help of (\ref{kw}),
and hence the statement provides an independent proof of (\ref{kw})
for the case $\g=\sln$.)

Being uniquely determined, these \furu s coincide with ring structures
that appear in other areas and possess the same basic properties.
Examples of such structures are provided by the truncated tensor
products of the quantum groups
{\sf U}$_q$(\sln) with deformation parameter $q=\exp(2\pi\ii/(k+N))$
\cite{Algs,pasa,fugp2,fuva2}, by the \lrc s for the so-called induction
product of Hecke \alg s at these roots of unity \cite{gowe,gowe2}, and
by the spaces of edge variables in fusion-RSOS models \cite{kuna}.
\vskip 2mm

The \slnk\ \furu s possess the following further properties:
\begin{itemize}
\ibox As a ring the \furu s are generated by
$\phi^{}_{\la_{(j)}}$ with $j=1,2,...\,,n-1$. Thus all generators
$\phi_\la$ can be expressed as polynomials over \zet\ in the
$n-1$ variables $x_j\equiv
\phi^{}_{\la_{(j)}}$. The explicit formula is similar \cite{gepn9}
to the so-called Giambelli formula \cite{HILl} for \sln\ tensor products;
e.g.\ for $n=2$ the polynomials are the Chebyshev polynomials of
the second kind. By combining the polynomials $\phi_\la=\phi_\la(x_l)$
with (\ref{7.1}), one can then obtain the full set of \furu s explicitly.
\ibox In agreement with (\ref{ll}), the structure
constants are majorized by the tensor product coefficients of \sln.
\ibox The fusion ring is isomorphic to the quotient of the tensor
product ring of \sln\ by the ideal that is generated by $\{\phi_\la\mid
\la\in I_{k+1}\!\setminus I_k\}$ \cite{gona}.
\ibox The fusion ring is also isomorphic to the quotient
$\zet[x_1,x_2,...\,,x_{n-1}]/{\cal J}$ of the ring
of polynomials in $n-1$ complex variables $x_j$
by the vanishing relations that are obtained when expressing the
\furu s of the $\phi^{}_{\la_{(j)}}$ entirely in the variables $x_l$.
\\The two-sided ideal $\cal J$ is generated by $\{\phi_{(k+l)\la_{(1)}}
^{}\mid l=1,2,...\,,n-1\}$ \cite{gepn9}.
\ibox The relations generating $\cal J$ can be integrated,
$\phi_{(k+l)\la_{(1)}}^{}(x)$ $=\partial V(x)/\partial x_l$, with potential
$V(x)=(x_1)^{n+k}+\ldots\;$. The full expression of $V$ in terms of the
variables $x_l$ is rather lengthy (for $n=2$, where there is a single
variable $x=x_1$, $V(x)$ is a Chebyshev polynomial of the first kind).
But when defining auxiliary variables $q_i$, $i=1,...\,,n-1$ by
  \be  x_i=\sum_{1\leq j_1<j_2<...<j_i\leq n} q_{j_1} q_{j_2} \ldots
  q_{j_i}, \ee
along with $q_n:=(\prod_{i=1}^{n-1}q_i)^{-1}$ (so that the Jacobian
$\partial x/\partial q$ of the transformation from $x$ to $q$ is the
Vandermonde determinant $\prod_{i<j}(q_i-q_j)$), the potential acquires
the simple form \cite{gepn9}
  \be  V(x)=V_{n,k}(x):=\frac1{n+k}\sum_{i=1}^n q_i^{n+k}. \ee
The expression in terms of the $x_l$ can be reduced, by means of a
$n+1\,$-term recurrence relation, to the result for low values of the
level. For instance, for $n=3$ there is the recurrence relation
  \be  V_{3,k}(x_1,x_2)=(k+3)^{-1} [ (k+2)\,x_1\,V_{3,k-1}-
  (k+1)\,x_2\,V_{3,k-2}+k\,V_{3,k-3}] \,,  \ee
by which one can compute $V_{3,k}$ from
$V_{3,0}=x_1^3/3-x_1^{}x_2^{}+1$,
$V_{3,1}=x_1^4/4-x_1^2x_2^{}+x_1^{}+x_2^2/2$, and
$V_{3,2}=x_1^5/5-x_1^3x_2^{}+x_1^{}x_2^2+x_1^2-x_2^{}$.
\ibox
The quasihomogeneous part of the potential $V(x)$ corresponds to
the ring of chiral primary fields of the $N=2$ superconformal coset
theory $({\sl sl}_{n+1})_k\oplus({\sl so}_{4n})_1/({\sl sl}_n)_{k+1}
\oplus u_1$, and also to the cohomology ring of the Grassmannian
manifold $U(n+k)/U(n)\times U(k)$ \cite{gepn9,intr2,bmrs}.
\ibox
For $n=3$, all eigenvalues of $\cn{\la_{(1)}}$ are distinct,
so that the fusion ring is polynomial with generator $x=x_1
\equiv\phi_{\la_{(1)}}$ \cite{dizu2}. The corresponding
polynomial is of order $4k-1$; e.g.\ one has
$V(x)=\frac17x^7-x^4-x$ at $k=2$, and
$V(x)=\frac1{11}x^{11}-\frac98x^8+\frac95x^5-4x^2$ at $k=3$.

\end{itemize}
 The \slnk\ fusion ring is particularly simple at level one.
In this case the index set is $I_1\cong\zet_n$, and the
fusion matrices read
  \be  (\cn i)_{jk}^{}=\delta_{i+j,k}^{(n)} ,   \labl{cz}
where $\delta_{ij}^{(n)}$ is 1 if $i=j\,{\rm mod}\,n$, and zero else.
Thus the \fra\ is isomorphic to $\complex\zet_n$. For any fusion algebra
with this property, the modular matrix $S$ has $n$th roots of unity
as its entries,
  \be  S_{jk}=n^{-1/2}\,\exp(2\pi\ii m\,jk/n) \,,  \ee
where $m$ is some integer coprime with $n$, whose value depends on the
precise identification of the fields with the elements of $\zet_n$.
Without loss of generality one can put $m=1$; for \sln\ at level one, this
corresponds to the natural
identification $j\in\zet_n\;\hat=\;\phi_{\Lambda_{(j)}}^{}$.

\subsect{Fusion graphs}

To any fusion matrix \cn i one may associate a labelled bicolored
graph \ga i\ and a labelled directed graph \gb i\ as follows
(see e.g.\ \cite{GOdj}).  To get \ga i, associate to any $j\in I$ a
`white' vertex $w_j$ and a `black' vertex $b_j$, and connect, for all
$j,k\in I$, $w_j$ with $b_k$ by $(\cn i)_{jk}$ edges.  Similarly, for
$i\neq i^+$, \gb i
is obtained by associating to any $j\in I$ a single vertex $v_j$ and
connecting, for all $j,k\in I$, $v_j$ with $v_k$ by $(\cn i)_{jk}$
directed edges; on the other hand, if $i=i^+$ so that \cn i is symmetric,
then for any edge from $j$ to $k$ this prescription would yield
precisely one edge from $k$ to
$j$; therefore in this case one simply connects $v_j$ and $v_k$ by
$(\cn i)_{jk}$ undirected edges. In short, \gb i is the graph whose
incidence (or connectivity, or adjacency) matrix is \cn i. Also note
that one can obtain \gb i\ from \ga i\ by
identifying $w_i$ with $b_i$, and by supplementing the edges between the
vertices with a direction `from white to black'.

These graphs serve as a compact description of the
fusion matrix. Moreover, a lot of structural information on the \fra\
can be read directly off the graphs. For instance, \cn i is
indecomposable iff \ga i is connected. Other examples are
\cite{hoff} certain properties of the Perron\hy Frobenius eigenvector,
as well as the following \cite{FRke}: For $\cd i>1$ a node in \gb i
corresponds to a primary field that is a simple current iff it is
reached by precisely one edge if $\phi_i$ is self-conjugate,
respectively by precisely two edges (one incoming and one outgoing)
if $\phi_i$ is non-selfconjugate.

If such a graph is isomorphic (as an unlabelled un-colored graph)
to the Dynkin
diagram of a simply laced simple or affine Lie algebra, the graph is
conventionally denoted by the name of this algebra. For the graph
corresponding to a simple Lie \alg\ \g,
the associated fusion rule matrix equals $2\,\one-$A, with A the
Cartan matrix of \g, and its eigenvalues are
$2\cos(\pi m_i/h)$, with $h$ the dual Coxeter number and $m_i$, $i=1,2,
...\,,{\rm rank}\,(\g)$, the exponents (i.e., the orders of
independent Casimir operators minus one) of \g. Similar abbreviations
as for the Dynkin diagrams can be used for other graphs, such as
  \be \begin{array}{lll}  {}\\[-4.4 mm] \bar A_r^{}&=&\picarbar,\\
  \bar A_{r-1}^{(1)} &=& \picaerbar,\\
  \bar D_{r-1}^{(1)} &=& \picderbar \\[-.4mm] {}\end{array}  \ee
(each of these has by definition $r$ nodes; thus, roughly,
$\bar A_r=\mbox{`}A_{2r}/\zet_2$',
$\bar A_r^{(1)}=\mbox{`}A_{2r+1}^{(1)}/\zet_2$', and
$\bar D_r^{(1)}=\mbox{`}D_{2r+1}^{(1)}/\zet_2$', where $\zet_2$ refers
to an appropriate automorphism of the relevant diagram).
If a graph is not connected, then the notation $\Gamma=\bigoplus_i\!
\Gamma_i$, is used, with $\Gamma_i$ the connected components.\\[3 mm]
Various examples of fusion graphs are given in the following list:
\begin{enumerate}\item
The simplest possibility for a pair of fusion graphs is given by
  \be  \gA=\mbox{$\bigoplus^\II\!$} A_2, \qquad\gB=
  \mbox{$\bigoplus^\II\!$} \bar A_1. \labl{g1}
It describes the fusion rules of the unit $\phi_0$; conversely,
$\phi_0$ is uniquely characterized by having (\ref{g1}) as its
fusion graphs.
\item
Numbering the canonical elements that span the \fra\ \czi\ according
to $\zet_\II$,
one has $\ga j$ as in (\ref{g1}) and $\gb j=\bigoplus^{\II/p}\!
A_{p-1}^{(1)}$ for $j=0,1, ...\,,\II$, where $p\geq2$ is
defined by setting $\Ii/j=p/q$ with $p$, $q$ coprime. For $j=0$, this
degenerates to (\ref{g1}), while for $j$ coprime to \Ii, the
graph is connected, $\gb j=A_{\II-1}^{(1)}$.
\item
One has $\ga j=\bigoplus^\II A_2$ iff $\phi_j$ is a simple current.
Also, if $\phi_j$ is a simple current, then
$\gb j=(\bar A_1)^{\oplus m}_{}\oplus (A_2)^{\oplus n}_{}\oplus
\bigoplus_j A_{l_j}^{(1)}$ for some integers satisfying $l_j\geq2$
and $m+2n+\sum_j(l_j+1)=\II$.
\item
Consider the fusion matrix
  \be  \cn1=\matrx{cc}{0&1\\[.5 mm]1&1} \,;  \labl{71}
this defines (along with, of course, $\cn0=\one$) the so-called
the Lee\hy $\!$Yang \furu s (these already appeared in (\ref{ff})
above). The associated graphs are $\ga1=A_4$ and $\gb1=\bar A_2$.
\\Similarly, for
  \be  \cn1=\matrx{ccc}{0&1&0\\[.5 mm]1&0&1\\[.5 mm]0&1&0} , \qquad
  \cn2=\matrx{ccc}{0&0&1\\[.5 mm]0&1&0\\[.5 mm]1&0&0} ,    \labl{72}
known as the Ising \furu s, the fusion graphs are
$\ga1=A_3\oplus A_3$, $\gb1=A_3$ and $\ga2=A_2\oplus A_2\oplus A_2$,
$\gb2=\bar A_1\oplus A_2$, respectively.
\item
For \slnk, $\gb{\Lambda_{(1)}}$ is a
graph obtained by `filling up' an $(n+1)$-gon of edge length $k+1$
in an appropriate manner; the corners correspond to the simple
currents. For $n=2$, one simply has $\gb{\Lambda_{(1)}}=A_{k+1}$.
\\For $n=3$, one has to fill the triangle of length $k+1$ with triangles
of length one, leading to $\gb{\Lambda_{(1)}}=A_2^{(1)}$ for $k=1$, and to
  \be  \begin{array}{c} \picslzkfg \\[-6 mm] {} \end{array}\labl{tria}
\vskip 10mm
for the general case. Here the labelling of most of the nodes has been
suppressed; it can easily be restored. In particular one has
  \be  \begin{array}{c} \picslzzfg \\[-6 mm] {} \end{array} \labl{zz}
\vskip 3mm
at level two.\\
For $n=4$, the situation is already a lot more complicated;
e.g.\ at level two, the graph $\gb{\Lambda_{(1)}}$ is given by
  \be  \begin{array}{c} \picsldzfg\\[-9 mm] {} \end{array} \ee
\vskip 11mm
\item The fusion graphs of the $(B_r)_2^{}$ \wzwt\ are:
$\gb{2\Lambda_{(1)}}=A_1^{(1)}\oplus A_1^{(1)}\oplus (\bar A_1^{})^
{\oplus r}$; $\gb{\Lambda_{(i)}}$, $i=1,2,...\,,r-1$, and
$\gb{2\Lambda_{(r)}}$ are all given by $\bar A_1^{(1)}\oplus\bar
D_{r+1}^{(1)}$; finally, $\gb{\Lambda_{(r)}}$ looks like
  \be  \begin{array}{c} 
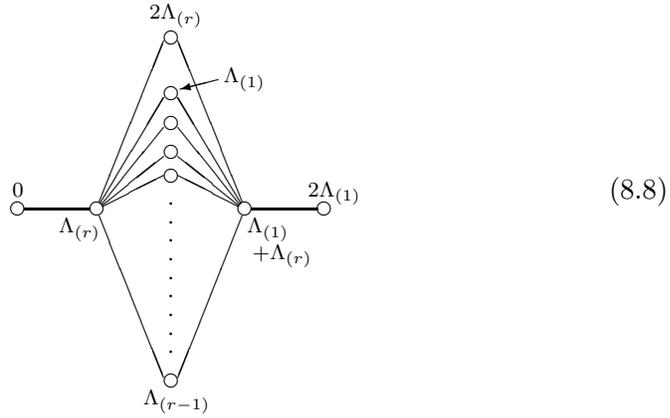
\begin{picture}(160,160)(0,-80)
 \putss{-1.8}{4.7}0            \putss{78}{48}{$\la_{(1)}$}
 \putss{110}{4.7}{$2\la_{(1)}$}\putss{15.6}{-9}{$\la_{(r)}$}
 \putss{87}{-9}{$\la_{(1)}$}   \putss{89}{-19}{$+\la_{(r)}$}
 \putss{50}{72}{$2\la_{(r)}$}  \putss{47.7}{-74}{$\la_{(r-1)}$}
 \mpcir00{30}025  \mpcir{86}0{30}025 \mplin{2.5}0{86}0210{25}
 \pucir{58}{-65}5 \pucir{58}{65}5  \pucir{58}{44}5 \pucir{58}{32.5}5
 \pucir{58}{21.5}5\pucir{58}{12.5}5 \mpcis{58}20{-7}91
 \pulin{30.9}{2.7}25{24.4}   \pulin{85.1}{2.7}{-2}5{24.4}
 \pulin{31.35}{2.2}35{24.05} \pulin{84.65}{2.2}{-3}5{24.05}
 \pulin{31.85}{1.7}45{23.45} \pulin{84.15}{1.7}{-4}5{23.45}
 \pulin{32.15}{1.4}54{23.15} \pulin{83.85}{1.4}{-5}4{23.15}
 \pulin{32.3}{1.1}21{23}     \pulin{83.7}{1.1}{-2}1{23}
 \pulin{30.9}{-2.7}2{-5}{24.4}\pulin{85.1}{-2.7}{-2}{-5}{24.4}
 \puvec{76}{49}{-4}{-1}{14.3}
 \end{picture}
  \\[-11 mm] {} \end{array}  \labl{8.6}
\vskip 4mm
and $\gb{\Lambda_{(1)}+\Lambda_{(r)}}$ is given by (\ref{8.6}) with
the labelling of the left- and rightmost nodes interchanged.
\item
Finally note that, of course, the fact that $\ga i=\ga j$ (as
unlabelled graphs) does not imply that $\gb i=\gb j$, too. In particular,
the connectivity of $\gb i$ is not fixed by $\ga i$. For instance,
for the $(A_2)^{}_2$ \wzwt, one has $\ga{\la_{(1)}}=A_4\oplus A_4
\oplus A_4= \ga{\la_{(1)}+\la_{(2)}}$, but
$\gb{\la_{(1)}}$ is given by the graph (\ref{zz}) above, whereas
$\gb{\la_{(1)}+\la_{(2)}}=A_2\oplus A_2\oplus A_2$.
\end{enumerate}

\noindent
Clearly, as soon as some \frc s become large, the fusion graphs
defined according to the above prescription become more and more
unhandy. A modest improvement is achieved by modifying the
prescription in the following manner. For self-conjugate $\phi_i$,
join the vertices $j$ and $k$ by a single edge iff $\nijk\neq0$,
and attach the number \nijk\ as a label to any edge with $\nijk>1$,
and similarly in the non-selfconjugate case.

\subsect{Classification}

A complete classification of \fra s (up to isomorphism)
is far beyond reach. This problem may well be as hard as, say,
the problem of classifying all discrete groups. In the modular
rational case, however, several
strategies which provide insight into the classification do exist.

\begin{enumerate}\item{Enumeration}\end{enumerate}
{}
The most direct approach is certainly
to translate, for a given index set $I$, the
various constraints provided by (F1) to (F4) and by (R)
into a parametrization of the most general solution.
This approach has been followed in
\cite{capr2} (see also \cite{capo2,rivl}).
The complexity of the system of equations grows rapidly with the
dimensionality \Ii\ of the algebra. A solution which exists for any
\Ii\ is given by \czi, as has already been observed in (\ref{cz}) above.
More generally, for any finite abelian group {\dl G}, \complex{\dl G}
is a rational fusion algebra.
An enumeration of all solutions is easy for $\Ii=2$ and $\Ii=3$, but
already for $\Ii=4$, this enumeration scheme becomes rather non-trivial
so that only some partial results \cite{capr2} are known.

For $\Ii=2$, besides $\complex\zet_2$ the only possibility is given by
the Lee\hy $\!$Yang \furu s (\ref{71}).
For $\Ii=3$, there are two solutions besides $\complex\zet_3$; one of
them describes the Ising \furu s (\ref{72}), and the other one is
  \be  \cn1=\matrx{ccc}{0&1&0\\[.5 mm]1&0&1\\[.5 mm]0&1&1} , \qquad
  \cn2=\matrx{ccc}{0&0&1\\[.5 mm]0&1&1\\[.5 mm]1&1&1}  .   \labl{73}
Both of these are polynomial \fra s: they satisfy $\phi_2^{}=P_2
(\phi_1)=\phi_1^2-1$, and the fusion potential is given by
$V(\phi_1) =\frac14\phi_1^4-\phi_1^2$ and
$V(\phi_1) =\frac14\phi_1^4-\frac13\phi_1^3-\phi_1^2+\phi_1^{}$,
respectively.
The fusion dimensions are $\cd1=\sqrt2$, $\cd2=1$ in the Ising case,
and $\cd1=2\,{\rm cos}(\pi/7)$, $\cd2=1+2\,{\rm cos}(2\pi/7)$
for (\ref{73}).

Many of the \fra s mentioned so far possess a realization as the \furu s
of some \wzwt. For example, according to section 7, a \wzwt\
realizing \czi\ is ({\sl sl}$_\II$)$^{}_1\equiv(A_{\II-1})
^{}_1$; the Lee\hy $\!$Yang \furu s are realized by
$(G_2)^{}_1$ and $(F_4)^{}_1$, and the Ising \furu s by
$(B_r)^{}_1$ for any rank $r$ and by $(E_8)^{}_2$.

Another approach to enumeration is to simply list the \fra s
that arise in certain classes of \cfts\ for which they are
calculable by some algorithm, say in \wzwts.
While this is usually not very illuminating, it can
nevertheless be helpful in specific circumstances.
For example, with this procedure one may encounter theories
with identical or closely related fusion rules, which in turn
suggests that other quantities for the corresponding \cfts\
(such as the modular matrix $S$ or correlation functions) are closely
related as well. Let me present a few examples. First,
as just mentioned, the \fra s of the \wzwts\
$(F_4)^{}_1$ and $(G_2)^{}_1$ coincide, and the same is true for
the $(B_r)^{}_1$ and $(E_8)^{}_2$ \wzwts. Similarly,
$(F_4)^{}_3$ has the same \furu s as $(G_2)^{}_4$, and
$(E_8)^{}_2$ the same as $(F_4)^{}_2$ \cite{fuva2}. There also
exists an infinite series of such correspondences, namely \cite{mnrs}
  \be  \ca((C_r)^{}_k) \cong \ca((C_k)^{}_r) \,; \ee
this is known as {\em level-rank duality.} Further, even more often
one arrives at identical structures after `modding out simple
currents', i.e.\ taking only one (specific) representative of
each simple current orbit (the physical interpretation of this modding
procedure still has to be clarified); for instance, one has
\cite{kuna,nasc2,fuva,nats}
  \be  \ca((A_r)^{}_k)/\zet_{r+1} \cong \ca((A_{k-1})^{}_{r+1})/\zet_k
  \,. \ee

The isomorphisms among fusion algebras of \wzwts\ translate to
similar relations for \cfts\ obtained from them via the coset construction.
In some cases they can be used to show that some a priori different
coset theories are actually identical
 \footnote{~At least modulo the ambiguities that can arise (see section
10 below) in the calculation of the \opa\ from the fusion rules.}
 as \cfts\ \cite{albs,fuSc2}.

\begin{itemize}\item[2.]{Enumeration of characters}\end{itemize}
{}
It can be shown that the conformal weights and conformal central charge
of a \rcft\ are rational numbers \cite{anmo}, implying that the modular
matrix $T$ obeys $T^m=\one$
for some integer $m$. Accordingly, it is expected that
for any \rcft\ the relevant \rep\ of the modular group factorizes through
a finite index normal subgroup $\Sigma$ of \pslz, i.e.\ the characters
are invariant under $\Sigma$ so that
the action of modular transformations on the characters generates the finite
group $\pslz/\Sigma$ \cite{hana,mase}. Accordingly, one may enumerate
the possible forms of the modular matrices $S$ and $T$ by first
making a list of all finite groups that possess a \rep\ of dimension
$M$ equal to the number of distinct characters of the theory (which
due to (\ref{ci}) is smaller than \Ii\ if some fields are
non-selfconjugate), then for each such group find the \rep\ matrix
for $S$, and then calculate the \furu s from the Verlinde formula
\cite{mase}. The first part of this programme is again involved
already for modestly large $m$ (the full list of relevant finite
groups is known only for $M=2$ and $M=3$ \cite{GRey}).
Further, for all \rcfts\ for which the characters are known explicitly,
one has in fact $\Sigma={\it SL}(2,\zet_m)$, so that one may employ
\cite{hana,ehol2}
the \rep\ theory of {\em SL}$(2,\zet_m)$ to describe the structure
of the associated \fra s.

\begin{itemize}\item[3.]{Subalgebras}\end{itemize}
{}
Given some \fra\ \ca, various further \fra s are provided by its \frsa s,
among which one may hope to find some that were not known previously.
In view of the poor prospects of the enumeration strategies,
this is not an extremely attractive approach.
But at least there exists a simple finite algorithm that provides all
\frsa s of a rational \fra\ \cite{FRke}. Namely, to any subset
$J\subseteq I$ one can associate a subset $J^\infty\subseteq I$ such
that $\{\phi_i\mid i\in J^\infty\}$ spans a \frsa,
and this exhausts the collection of \frsa s of \ca. The set
$J^\infty$ is determined as follows \cite{FRke}. For
$J\subseteq I$ define $J^+:=\{j\in I\mid j^+\in J\}$, and for any
pair of subsets
$J,J'\subseteq I$ define $J\star J':=\{j''\in I\mid \sum_{j\in J,
j'\in J'}{\cal N}_{jj'}^{\ \,j''}\neq0\}$; write $J\star J$ as
$J^2$, and recursively $J^m:=J^{m-1}\star J$. Then
  \be  J^\infty=\bigcup_{m,n\in\zetplusO}J^m\star(J^+)^n. \ee

A special example of this construction is given by the set
  \be  \Jo=\{j\in I\mid \cd j=1\}  \ee
of simple currents. By elementary properties of simple currents,
one has $\Jo^+=\Jo$ and $\Jo^2=\Jo$, and hence $\Jo^\infty=\Jo$. The
associated \frsa\ is simply $\complex \Jo$, with $\Jo$ regarded as an
abelian group (with the multiplication $i\star j$ induced by the
fusion product, and the inverse $i^{-1}=i^+$ induced by conjugation).
In particular, $\Jo\cong\zet_{|\Jo|}$ if $|\Jo|$ is not divisible by a square
number.

Another simple example of subalgebras arises in connection with
the {\em tensor product\/} of \fra s. By definition, the tensor
product (also called crossed product in \cite{FRke}) $\ca={\cal A}_1
\times{\cal A}_2$ of two \fra s ${\cal A}_1$ and ${\cal A}_2$ with
index sets $I_1$ and $I_2$ is the algebra with canonical basis
  \be  \{\Phi_{ij}\mid i\in I_1,\;j\in I_2\}=:\{(i\otimes j)\} \,, \ee
and product
  \be  (i\otimes j)\star(i'\otimes j')=((i\star_{\scriptscriptstyle1}i')
  \otimes(j\star_{\scriptscriptstyle2}j')) \,. \ee
Clearly, this describes again a \fra, with unit $\bfe=(0\otimes0)$
and conjugation $(i\otimes j)^+_{}=(i_{}^{+_1}\otimes j_{}^{+_2})$,
and ${\cal A}_1$ and ${\cal A}_2$ are \frsa s of \ca.

One may identify the sets $I_1$ with $\{(i\otimes0)\mid i\in I_1\}$ and
$I_2$ with $\{(0\otimes j)\mid j\in I_2\}$, thereby considering them
as subsets of the index set $I$ of \ca; they are then characterized by
  \be  I_1^+=I_1=I_1\star I_1, \qquad I_2^+=I_2=I_2\star I_2, \qquad
  I_1\cap I_2=\{0\}, \ee
and by the property that any $i\in I\setminus(I_1\cup I_2)$ obeys
  \be  \{i\}=\{i_1\}\star\{i_2\}   \ee
for a uniquely determined pair $i_1\in I_1$, $i_2\in I_2$ of indices.
Conversely, any \fra\ whose index set $I$ possesses subsets $I_1$
and $I_2$ with these properties is a tensor product of smaller \fra s,
and hence need not be considered separately in the classification
programme.

\begin{itemize}\item[4.]{Grading} \end{itemize}
{}
A \fra\ \ca\ is said to be $\zet_n$-graded iff there exists a partition
  \be  I=\bigcup_{p\in\zet_n}\!K_p   \ee
of the index set $I$ that satisfies
  \be  K_0\ni0, \qquad K_{-p}=K_p^+, \qquad {\rm and} \qquad
  K_p\star K_q=K_{p+q} \ee
for all $p,q\in\zet_n$; the integer $p\in\zet_n$ is called the grading
of the elements $\phi_i$ for which $i\in K_p$. (Below it will usually be
assumed that the maximal possible grading of \ca\ has been chosen; in
the trivial case $n=1$, i.e. $I=K_0$, one calls \ca\ ungraded). To any
$j\in I$, one can associate a $\zet_{n_j}^{}$-gradation of the \frsa\
of \ca\ that corresponds to the index set $\{j\}_{}^\infty$; this
gradation is uniquely determined by the property that $K_0=(\{j\}\star
\{j^+\})_{}^\infty$ (which implies that $n_j\in\{1,2\}$ if $j$ is
self-conjugate).

Given a $\zet_n$-gradation, the set $\Jo$ of simple currents obeys
  \be  \Jo/(\Jo\cap K_0)\cong\zet_m  \ee
for some $m\in\zetplus$ that is a common divisor of $n$ and $|\Jo|$.
It can be shown \cite{FRke} that the entire algebra \ca\ is already
determined by its restriction to $\bigcup_{p\in\zet_m}\!K_p$, supplemented
by the \furu s of a specific simple current that belongs to $K_m$.

The grading of a \fra\ becomes relevant to the classification problem
because of the following two constructions \cite{FRke} of new \fra s
from known ones. To any \fra\ \ca\ one can
associate for any $m\in\zetplus$ a \fra\ ${\cal A}^{[m]}$, and for
any $j_\circ\in\Jo(\ca)$ a \fra\ ${\cal A}^{\{j_\circ\}}$. A canonical
basis of ${\cal A}^{[m]}$ is
  \be  \{(\phi_i,p)\mid i\in I,\;p\in\zet_m\} \,, \ee
while the canonical basis of
${\cal A}^{\{j_\circ\}}$ coincides with the canonical basis of \ca.
The fusion product of ${\cal A}^{[m]}$ is defined by
  \be  (\phi_i,p)\star_{\scriptscriptstyle[m]}(\phi_j,q)
  =(\phi_i\star\phi_j,p+q+r_{ij}) \,, \ee
and the fusion product of ${\cal A}^{\{j_\circ\}}$ by
  \be  \phi_i\star_{\scriptscriptstyle\{j_\circ\}}\phi_j
  =(\phi_{j_\circ})^{r_{ij}}_{} \star\phi_i\star\phi_j \ee
(and similarly for the conjugation), where $r_{ij}$ is an integer
determined by the grading of $\phi_i$ and $\phi_j$.
Of course, these procedures can be iterated; one has
  \be  ({\cal A}^{\{j_\circ\}})^{\{k_\circ\}}
  \cong{\cal A}^{\{j_\circ\star k_\circ\}}, \quad
  ({\cal A}^{[m]})^{[n]}\cong{\cal A}^{[mn]}, \quad
  ({\cal A}^{\{j_\circ\}})^{[m]}\cong({\cal A}^{[m]})
  ^{\{(j_\circ)^m\}}. \ee

In addition to generating new \fra s from known ones, these constructions
allow conversely to reduce the classification of all \fra s to the
classification of those with certain specific properties. For instance
\cite{FRke}, it is sufficient to restrict to even-graded \fra s,
because any odd-graded algebra $\tilde{\cal A}$ appears in the list
of even-graded ones as $\tilde{\cal A}^{[2]}_{}=\ca$ with \ca\ such that
there exists a $j_\circ\in\Jo(\ca)$ with $j_\circ\star j_\circ=0$ and
$j_\circ\not\in K_0$.

\begin{itemize}\item[5.]{Polynomial \fra s}\end{itemize}
{}
Sometimes progress can be made by restricting to a subclass of \fra s
that satisfy some further axiom. A class for which some rather general
results are available is given by the polynomial \fra s satisfying the
polynomiality condition (\ref{84}) \cite{capr2}.
Note that a \fra\ is polynomial iff $\{j\}_{}^\infty=I$ for some $j\in I$.
Also, if \ca\ is polynomial with generator $\phi=\phi_1$, the fusion
graph $\Gamma(\cn1)$ is connected.

Restricting to self-conjugate generator $\phi_1$, all \furu\
eigenvalues $\nv j1$ are real and distinct. Therefore the polynomials
$p_i$ in (\ref{84}), considered as
functions over the real numbers, are orthogonal with respect
to some positive definite measure, and hence satisfy a three-term
recurrence relation. For the fusion matrices this implies that
they are tridiagonal, i.e.\ with appropriate labelling they obey
$(\cn i)_{jk}=0$ for $|i-j|\geq2$.
It then follows that the full information about a self-conjugate
polynomial \fra\ is contained in the $2\II-3$ integers
${\cal N}_{1ii}$ with $i\in\{1,...\,,\II-1\}$, and
${\cal N}_{1\,i\,i+1}$ with $i\in\{2,...\,,\II-1\}$.
These are still further restricted by associativity and by the recurrence
relation just mentioned. If all of them are zero or one, then
a complete classification has been obtained \cite{capr2}, while for
the general case there are only partial results.

\begin{itemize}\item[6.]{Small fusion dimensions.}\end{itemize}
{}
A complete classification is available for (not necessarily rational)
polynomial \fra s whose
generator $\phi=\phi_1$ has fusion dimension $\cd1\leq2$. It is based
on the important mathematical result \cite{GOdj} that the only connected
finite bicolorable graphs whose incidence matrices have Perron\hy
Frobenius eigenvalue smaller than two, are
  \be  A_r, \quad D_r, \quad E_r , \ee
i.e.\ the Dynkin diagrams of the simply-laced simple Lie algebras,
and that those for which the Perron\hy Fro\-be\-nius eigenvalue equals
two, are
  \be  A^{(1)}_r,\ r\geq2, \quad D^{(1)}_r, \quad E^{(1)}_r , \ee
i.e.\ the Dynkin diagrams of the simply-laced non-simple affine
Lie algebras except $A^{(1)}_1$.

Using this result, the following classification of polynomial \fra s
\ca\ with generator $\phi_1$ has been obtained in \cite{FRke}:
\begin{itemize}
\ibox If \ca\ is $\zet_2$-graded, and $\phi_1$ is self-conjugate with
fusion dimension $\cd1<2$, then $\Gamma(\cn1)$ is one of the Dynkin
diagrams
  \be  A_r,\ r\geq2, \quad D_{2r}, \quad E_6, \quad E_8, \labl{6-8}
and each of these possibilities corresponds to precisely one \fra.\\
For $\Gamma(\cn1)=A_r$, the fusion rules are those of the $A_1$
\wzwt\ at level $r-1$, with $\phi_1$ corresponding to the defining
representation of $A_1$. In particular, the conjugation is trivial,
$\Jo\cong\zet_2$, and $\cd1=2\,\cos(\pi/(r+1))$.\\
For $\Gamma(\cn1)=D_{2r}$, one has $\cd1=2\,\cos(\pi/(4r-2))$ and
$\Jo\cong\zet_3$ for $r=2$, $\Jo=\{\bfe\}$ else; the conjugation
is trivial for odd $r$, while for even $r$ it interchanges the
generators that correspond to the end points of the two short legs
of the diagram $D_{2r}$.\\
For $\Gamma(\cn1)=E_6$, one has $\cd1=2\,\cos(\pi/12)=(\sqrt3+1)/\sqrt2$
and $\Jo\cong\zet_2$, and the conjugation is trivial. Finally,
for $\Gamma(\cn1)=E_8$, one has $\cd1=2\,\cos(\pi/30)=\sqrt3(\sqrt5+1)
+\sqrt2\,\sqrt{5-\sqrt5}\,$ and $\Jo=\{\bfe\}$, and the conjugation is
again trivial.\\
($E_7$ and $D_{2r+1}$ do not appear in the list (\ref{6-8}). It can be
shown directly that such graphs do not provide a consistent \fra;
for instance, having $\Gamma({\cal N})=E_7$ would imply that one of
the generators would have fusion dimension $\,2\cos(5\pi/18)
\simeq1.28$, which is not contained in the allowed range (\ref H)
\cite{ocne,izum}.)
\ibox If \ca\ is $\zet_2$-graded, and $\phi_1$ is non-selfconjugate with
fusion dimension $\cd1<2$, then the \fra\ is one of
  \be  (A_{2r+1})_{}^{\{j_\circ\}}, \ r\geq2, \quad
  (E_6)_{}^{\{j_\circ\}},  \ee
where $A_{2r+1}$ and $E_6$ stand for the \fra s that in the manner
just described are associated to the respective Dynkin diagrams, and
where $j_\circ$ corresponds to the non-trivial simple current of these
algebras.
\ibox If \ca\ is ungraded, and $\cd1<2$, then \ca\ is the \frsa\ spanned
by the 0-graded generators of the \fra\ corresponding to the $A_{2r}$
Dynkin diagram for some $r\geq2$. The associated fusion graph is
  \be  \bar A_r, \ r\geq2,  \ee
the fusion dimension is $\cd1=2\,\cos(\pi/(2r+1))$, and the
conjugation is trivial.
\ibox If \ca\ is $\zet_2$-graded, and $\phi_1$ is self-conjugate with
fusion dimension $\cd1=2$, then $\Gamma(\cn1)$ is one of the
Dynkin diagrams
  \be  A^{}_\infty, \quad D^{}_\infty, \quad D^{(1)}_r, \quad E^{(1)}_r ; \ee
for $D^{(1)}_r$, there exist precisely two inequivalent \fra s,
while for the other possibilities the \fra\ is determined uniquely.
For each of these algebras, all fusion dimensions are integral.\\
The fusion ring corresponding to the diagram $A_\infty$ (i.e.\ the graph
obtained from the Dynkin diagram $A_r$ in the limit $r\rightarrow\infty$)
is the fusion ring of the $A_1$ \wzwt\ in the limit of infinite
level, i.e.\ the representation ring of the $A_1$ Lie algebra.
In particular, the fusion dimension ist just equal to the ordinary
dimension of the relevant $A_1$-representation.\\
For $\Gamma(\cn1)=D_\infty$ (i.e.\ the graph obtained from the Dynkin
diagram $D_r$ by extending the long leg infinitely), there is one
simple current besides \bfe, and all other generators have ${\cal D}=2$.
For the two algebras with $\Gamma(\cn1)=D_r^{(1)}$, one has $\Jo
\cong\zet_4$ and $\Jo\cong\zet_2\times\zet_2$, respectively, and
again all remaining generators have fusion dimension 2.\\
For $\Gamma(\cn1)=E_6^{(1)}$, the fusion dimensions are as displayed
in the following picture:
  \be \begin{array}{l} {}\\[4 mm] \picesixe \end{array}\ee
\vskip -6mm
The \furu s of the generators with ${\cal D}\in\{1,2\}$ follow
from the $\zet_3$-grading of the algebra that corresponds to the
$\zet_3$-symmetry of the $E_6$ diagram, while for the field
$\phi_3$ with fusion dimension 3, $\phi_3\star\phi_3$
contains each of the simple currents once and $\phi_3$ twice.
\ibox If \ca\ is $\zet_2$-graded, and $\phi_1$ is non-selfconjugate with
fusion dimension $\cd1=2$, then $\Gamma(\cn1)$ is one of
  \be  E_6^{(1)}, \quad (E_7^{(1)})_{}^{\{j_\circ\}}, \quad
  (D_r^{(1)})_{}^{\{j_\circ\}},  \ee
and if \ca\ is ungraded, with $\phi_1$ non-selfconjugate and
$\cd1=2$, then $\Gamma(\cn1)$ is one of
  \be  \bar D_r^{(1)}, \ r\geq3 .  \labl)
Each of these corresponds to a unique \fra. The algebra
with $\Gamma(\cn1)=\bar D_r$ is the \frsa\ spanned by the
0-graded generators of one of the two \alg s corresponding to the graph
$D^{(1)}_{2r-1}$.
\ibox Finally, if $\cd1\leq2$ and \ca\ is $\zet_n$-graded with $n>2$,
then there is a rather long list of
possibilities for $\Gamma(\cn1)$ \cite[Theorem\,3.4.11]{FRke}. All
of the corresponding \fra s can be obtained with the help of the
constructions $\ca\mapsto\ca_{}^{[m]}$ and
$\ca\mapsto\ca_{}^{\{j_\circ\}}$ that were described above from
the algebras appearing in the previous classifications (\ref{6-8})
to (\ref)), supplemented by two other series of algebras. One of the
latter series corresponds to the graphs
  \be  \begin{array}{l} {}\\[-32 mm] 
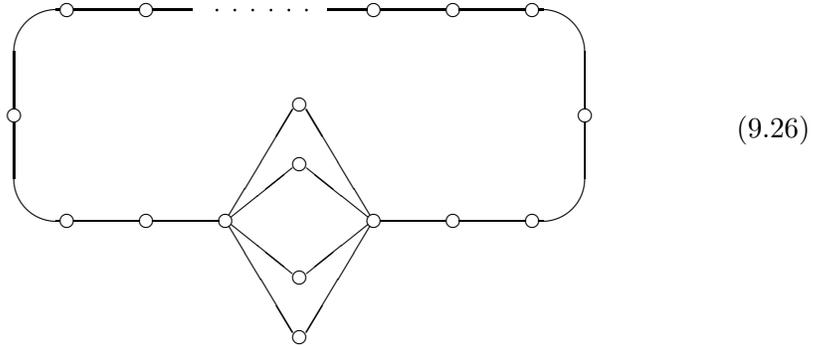
\begin{picture}(170,200)(0,-25)
 \mpcir{-30}0{30}035  \mpcir{86}0{30}035
 \mpcir{-30}{80}{30}025  \mpcir{86}{80}{30}035
 \mpcir{-50}{40}{216}025 \mpcis{26.8}{80}{6.8}061
 \mplin{-27.5}0{30}0210{25} \mplin{88.5}0{30}0210{25}
 \pulin{-27.5}{80}10{25} \mplin{88.5}{80}{30}0210{25}
 \mplin{2.5}{80}{66}0210{15}
 \pucir{58}{44}5 \pucir{58}{21.5}5 \pucir{58}{-44}5\pucir{58}{-21.5}5
 \pulin{31.35}{2.2}35{24.05} \pulin{84.65}{2.2}{-3}5{24.05}
 \pulin{32.15}{1.4}54{23.15} \pulin{83.85}{1.4}{-5}4{23.15}
 \pulin{31.35}{-2.2}3{-5}{24.05} \pulin{84.65}{-2.2}{-3}{-5}{24.05}
 \pulin{32.15}{-1.4}5{-4}{23.15} \pulin{83.85}{-1.4}{-5}{-4}{23.15}
 \put(-32.5,37.5){\oval(35,75)[bl]}\put(148.5,37.5){\oval(35,75)[br]}
 \put(-32.5,42.5){\oval(35,75)[tl]}\put(148.5,42.5){\oval(35,75)[tr]}
 \end{picture} \end{array} \ee
\vskip 8mm
while the graphs for the other series look much more complicated.
\end{itemize}

\noindent
Non-polynomial \fra s are not directly accessible to the methods of
\cite{FRke}. However, for \wzwts\ the list of all primary fields
with ${\cal D}\leq2$ is known \cite{jf17,jf18}, so that one can
identify the corresponding \fra s by inspection. The fusion graphs
for \wzw\ primaries $\phi_\la$ with fusion dimension
${\cal D}=2\cos(\pi/m)$ with $m\in\zet_{\geq4}$
are as displayed in the following table:
  \be  \begin{tabular}{|l|r|r|r||c|} \hline &&&&\\[-3.3 mm]
  \multicolumn{1}{|c|}{\g} & \multicolumn{1}{c|}{$k$}
  & \multicolumn{1}{c|}{$\la$} & \multicolumn{1}{c|}{$m$}
  & \multicolumn{1}{c|}{$\gb\la$} \\[-3.3 mm] &&&&\\
  \hline\hline &&&&\\[-3.3 mm]
  $A_{r-1}$ & 2 &&& \\
  $C_r$     & 1 & $\la_{(1)}^{}$ & $r+2$ & $A_{r+1}$ \\
  $A_1$     &$r$&&& \\
[-3.3 mm] &&&& \\ \hline &&&&\\[-3.3 mm]
  $B_r$     & 1 & $\la_{(r)}^{}$ && \\ [-1.8 mm] &&& 4 & $A_3$ \\[-1.8 mm]
  $E_8$     & 2 & $\la_{(1)}^{}$ && \\
[-3.3 mm] &&&& \\ \hline &&&&\\[-3.3 mm]
  $F_4$     & 1 & $\la_{(4)}^{}$ && \\ [-1.8 mm] &&& 5 & $\bar A_2$ \\[-1.8 mm]
  $G_2$     & 1 & $\la_{(2)}^{}$ && \\
[-3.3 mm] &&&& \\ \hline &&&&\\[-3.3 mm]
  $G_2$     & 2 & $\la_{(1)}^{}$ & 9 & $\bar A_4$ \\
[-3.3 mm] &&&& \\ \hline &&&&\\[-3.3 mm]
  $F_4$     & 2 & $\la_{(1)}^{}$ && \\ [-1.8 mm] &&&11 & $\bar A_5$ \\[-1.8 mm]
  $E_8$     & 3 & $\la_{(8)}^{}$ && \\
[-3.3 mm] &&&& \\ \hline &&&&\\[-3.3 mm]
  $E_7$     & 2 & $\la_{(7)}^{}$ & 4 & $A_3\oplus A_3$ \\
[-3.3 mm] &&&& \\ \hline &&&&\\[-3.3 mm]
  $E_7$     & 2 & $\la_{(1)}^{}$ & 5 & $\bar A_2\oplus\bar A_2\oplus\bar A_2$
\\
[-3.3 mm] &&&& \\ \hline &&&&\\[-3.3 mm]
  $E_6$     & 2 & $\la_{(2)}^{}$ & 7 & ${\cal E}_9$ \\
[-3.3 mm] &&&& \\ \hline \end{tabular}\labl{TA}
 \vskip 3mm
Note that if $\gb\la$ is connected, then the \fra\ is polynomial,
and hence is one of the \alg s described in the previous list.
For theories containing simple currents, only one representative
$\phi_\la$ of a simple current orbit is written in column 3 of the
table (\ref{TA}); also, the graph appearing in the last line is
  \be  \begin{array}{lll} {}\\[-19 mm] {\cal E}_9&=& \piccalenine
  \\[11 mm] {}\end{array}\ee
\vskip 3mm

The fusion graphs of \wzw\ primaries with ${\cal D}=2$ are shown in
the next table -- with the exception of the fields $\phi^{}_{\la_{(j)}}$,
with $j=2,3,...\,,[r/2]$, of the $(D_r)_2^{}$ theory. For the latter,
the fusion graphs are disconnected,
with connected components of the type $A_r^{(1)}$, $D_r^{(1)}$,
$\bar A_r^{(1)}$, and\,/\,or $\bar D_r^{(1)}$, but the systematics
of the decomposition into these connected components is somewhat
complicated.
  \be  \begin{tabular}{|l|r|l||c|} \hline &&&\\[-3.3 mm]
  \multicolumn{1}{|c|}{\g} & \multicolumn{1}{c|}{$k$}
  & \multicolumn{1}{c|}{$\la$} & \multicolumn{1}{c|}{$\gb\la$}
  \\[-3.3 mm] &&&\\ \hline\hline &&&\\[-3.3 mm]
  $C_4$     & 1 & $\la_{(2)}^{}$  & \\ [-1.8 mm] &&&
              $\bar A_1^{(1)}\oplus\bar D_2^{(1)}$ \\[-1.8 mm]
  $A_1$     & 4 & $2\la_{(1)}^{}$ & \\
[-3.3 mm] &&& \\ \hline &&&\\[-3.3 mm]
  $A_2$     & 3 & $\la_{(1)}^{},\;\la_{(2)}^{}$ & ${\cal T}_{10}$ \\
[-3.3 mm] &&& \\ \hline &&&\\[-3.3 mm]
  $A_3$     & 2 & $\la_{(2)}^{}$ & $A_3^{(1)}\oplus D_5^{(1)}$ \\
[-3.3 mm] &&& \\ \hline &&&\\[-3.3 mm]
  $B_r$     & 2 & $\la_{(j)}^{},\,{\scriptstyle j=1,2,...\,,r-1;}\;
         2\la_{(r)}^{}$ & $\bar A_1^{(1)}\oplus\bar D_{r+1}^{(1)}$ \\
[-3.3 mm] &&& \\ \hline &&&\\[-3.3 mm]
  $D_4$     & 2 & $\la_{(3)}^{},\;\la_{(4)}^{}$
            & $A_3^{(1)}\oplus D_6^{(1)}$ \\
[-3.3 mm] &&& \\ \hline &&&\\[-3.3 mm]
  $D_r$     & 2 & $\la_{(1)}^{}$ & $A_3^{(1)}\oplus D_{r+2}^{(1)}$ \\
[-3.3 mm] &&& \\ \hline \end{tabular}\ee
 \vskip 2mm \noindent
Here ${\cal T}_{10}$ stands for a triangular array of the type
encountered in (\ref{tria}) above.

\begin{itemize}\item[7.]{Local algebras.}\end{itemize}
{}
According to the results of section 6 any modular rational
\fra\ can be viewed as the local algebra of some polynomial
$V$. Thus one may try to classify \furu s to some extent
by classifying local algebras.
In particular, one can start from any quasihomogeneous polynomial
and deform it in such a way as to obtain a fusion algebra.
However, the conditions for a deformation to provide a \fra\
(that is, essentially, to allow for a conjugation that is unital)
are rather non-trivial and to my knowledge no general algorithm for
obtaining solutions is known so far. On the other hand,
the classification of quasihomogeneous polynomials (with isolated
singularities) can be achieved for any fixed number $n$ of
variables, and indeed has been completed \cite{krsk,klsc})
for small $n$.

Note that even if one succeeded in classifying some class of
allowed fusion potentials, one still has to investigate the
equivalences
among them, since as seen in section 6 a given fusion algebra can
possess many distinct presentations as a local algebra.

\subsect{The conformal bootstrap}

Given some physical realization of a \cft, the
\opc s of the theory are quantities which can, in principle, be
measured experimentally. For example, if the \cft\ is used in
the inner sector of a string compactification, then the
the Yukawa couplings of the massless particles
that are present in the low energy limit are (products of) \opc s
(compare e.g.\ \cite{sost,kaki,grlr,scsc}). Similarly, if the theory
describes a statistical system at a second order phase transition,
the \opc s determine the corrections to finite size scaling \cite{card3}.
Furthermore, if the \opc s as well as the symmetry algebra of a
\cft\ are known, the theory can be considered as completely
solved. For all these reasons, it is of considerable interest
to compute the \opc s of a theory. It is the central idea of the
conformal bootstrap programme to perform this calculation on the
basis of a few general requirements, among which the prominent part
is played by the associativity of the \opa.

By their definition, the \furu s of a \cft\ carry an important amount
of information about the structure of the \opa. However,
as mentioned in section 9, {\em many\/} different \cfts\ may possess
the same \fra. For example, the
$(B_r)^{}_1$ \wzwts\ for any rank $r$, and also the $(E_8)^{}_2$
theory, all realize the Ising \furu s; and even the trivial
one-dimensional \fra\ described by $\bfe\star\bfe=\bfe$ is shared
\cite{sche5} by a huge class of unitary modular invariant theories.
 But still, the \furu s of a \twodim~\cft\ at least provide nontrivial
constraints that allow for a partial solution of the bootstrap
programme. To explain this, I will now describe how far one can get in
determining the operator products by using the \furu s as an input.  As
mentioned towards the end of section 3, given the \four s of all primary
fields, one can determine the \opc s and hence solve the theory
completely.  The \four s, in turn, are to a large extent fixed by the
analytic properties of their chiral blocks,
and (part of) these analytic properties can be deduced from the \furu s.

What is relevant are actually not the chiral blocks themselves, but
rather how they combine, according to (\ref{14}), to the full
correlation functions. The information carried by the coefficients
in (\ref{14}) is equivalent to the information contained in the so-called
fusing and braiding matrices $F={\sf F}\jkil$
and $B={\sf B}\jkil$ which, according to
  \be  {\cal F}_{ilkj,p}(z) ={\displaystyle\sum_{m}}\,{\sf F}_{pm}
  \jkil\,{\cal F}_{ijkl,m}(z^{-1}) \quad{} \labl{fb}
and
  \be  \,{\cal F}_{ikjl,n}(z) ={\displaystyle\sum_{m}}\, {\sf B}_{nm}
  \jkil\, {\cal F}_{ijkl,m}(1-z) ,  \labl{bf}
implement the duality transformations
`fusing' and `braiding' on the space of chiral blocks. Note that
the chiral blocks may be represented pictorially as
in the figure (\ref G), where the external lines are to be interpreted as
the primary fields in $\cal F$, and the internal lines as the families
that are exchanged in the $s$-, $t$-, and $u$-channel, respectively;
in this picture, the fusing and braiding matrices relate the
chiral blocks of the $s$-channel to those of the $u$- and $t$-channel,
respectively.

There exist two different strategies for determining the matrices $F$ and
$B$.  One possibility \cite{blya2} is to solve a set of consistency
conditions known as (genus zero) polynomial equations which can be
deduced \cite{mose} by applying duality transformations to five-point
functions.  The first of these relations, the pentagon equation,
describes the compatibility of fusing and braiding; the second, the
hexagon equation, corresponds to the \ybe\ for the \rep\ of the braid
group that is induced by $B$.  In the second approach \cite{jf14},
linear \deq s for the correlators are constructed (often these also
follow from the presence of null vectors in the Verma modules of $\cal
W$; an example are the \kze s \cite{knza} for \wzw\ correlators).  Their
independent solutions are  the chiral blocks, and $B$ and $F$ can be
determined from the explicit form of the blocks.

The \furu s come into this game as follows.  In order to know the
dimensionality of the duality matrices, respectively the order of the
relevant differential equation, one must know the number of chiral blocks
that contribute to the four-point correlators.  According to the
formula (\ref M), this information can be read off the \furu s.

To be able to assess the strength of these ideas, it will be necessary to
mention a few details \cite{jf14,jf21} of the second strategy. Thus
consider
the \four\ (\ref{11}).  From the structure of the \opa\ (\ref o), it
follows that the chiral blocks in the $s$-channel behave as
$z^{-\Delta_i-\Delta_j+ \tilde\Delta_m^{(ij)}}$ for $z\rightarrow 0$,
where $\tilde\Delta_m^{(ij)}=\Delta_m^{(ij)} + \mu_m^{(ij)}$ is the
\cdim~of that field $\varphi_m$, of grade $\mu_m^{(ij)}$,
in the family $[\phi_m]\equiv[\phi_m^{(ij)}]$
that is responsible for the leading contribution to the coupling
between $\phi_i$, $\phi_j$ and $[\phi_m]$.
 \footnote{~The notation used here applies directly only to the case
${\cal N}_{ij}^{\ \,m}\leq1$; otherwise a multiplicity index
distinguishing the ${\cal N}_{ij}^{\ \,m}$ possible couplings is
needed.}
 Similarly, in the $t$- and $u$-channel the singularities are of the
form $(1-z)^{-\Delta_i-\Delta_k+ \tilde\Delta_n^{(ik)}}$ for
$z\rightarrow 1$, and $(z^{-1})^{\Delta_i-\Delta_l+
\tilde\Delta_p^{(il)}}$ for $z\rightarrow\infty$, respectively.  The
systems of chiral blocks in the three channels are not
independent. Associativity of the \opa~implies
$\calf_{ijkl}(z,\bar{z})= \calf_{ikjl}(1-z,1-\bar{z})=
z^{-2\Delta_{i}}\bar{z}^{-2\bar{\Delta}_{i}}\, \calf_{ilkj}\!
\left(z^{-1},\bar z^{-1}\right)$, which in turn requires
\cite{lewe} that the systems are linearly related through
analytic continuation; this is the contents of the relations
(\ref{fb}) and
(\ref{bf}). Moreover, in each channel the system of chiral blocks must
be \alg ically independent, which implies that the Wronskian
determinant of the system must not vanish identically.  Putting this
information together, it follows from elementary results of the theory of
ordinary linear \deq s that the chiral blocks are the $M$ independent
solutions of an $M$th order \deq~in the variable $z$, with only regular
singular points, among which there are in particular $z=0,\,1$ and
$\infty.$ The general solution of this equation is described by a
so-called Riemann scheme which specifies the positions of the
singularities together with the exponents \aim~at $z_i$; the latter are
the roots of the the $M$th order \alg ic equation that one obtains in
lowest order in $z-z_i$ by inserting the ansatz
$\calf(z)=(z-z_i)^{\alpha^{(i)}}\sum_{p=0}^\infty a_p(z-z_i)^p\,$
($a_0\neq0$) into the \deq.  The Riemann scheme does not, in general,
determine the chiral blocks uniquely because the number of parameters on
which the \deq, and hence its solutions, depend is generically larger
than the number of exponents (the additional parameters needed to specify
the \deq\ uniquely are called accessory parameters).

The ideas described above can for example be employed to gain
insight into the classification of (quasi)\rcfts. Namely, suppose
that the singularity structure of the \four s is
fully known, which means that the \cdim s of the primaries,
including the integer part (and, in the case of \frc s larger than
1, also the grades of the exchanged fields which are responsible for
the leading singular behaviour of a chiral block) are given.
Then it can be proved \cite{blya2} that the duality matrices are
uniquely determined, provided that the \deq s satisfied by the
\four s do not possess any \asis. (By definition, an \asi~of a
\deq~is a singular point of the equation at which any of its
solutions is regular.) In particular, in the absence of \asis\ the
values of all
accessory parameters can be fixed by imposing the \poleq s.
On the other hand, the presence of \asis~spoils this uniqueness
property, because the local monodromy of the solutions around an \asi~is
trivial so that the number and positions of \asis~appear as free
parameters in the Riemann monodromy problem.

Note that only the positions and exponents of the real singularities
of correlation functions are part of the basic data of a \cft, while
the number and positions of \asis\ must be considered as arbitrary, up to
mild restrictions which result \cite{jf21} from
crossing symmetry. (The \poleq s, including those arising at
genus one which involve the modular transformation matrices $S$ and $T$,
do not lead to any further constraints on the \asis; this is so
because duality is the `square root' of monodromy \cite{mose};
technically, it follows by application of
a simple result from the theory of isomonodromic deformations of \deq s.)
 If the \cdim s are only known up to integers,
one can consider the integer parts, and similarly also
the grades $\mu_m^{(i\,\cdot)}$ mentioned above,
as additional parameters of the classification programme, and perform the
analysis for each allowed set of
parameters separately. One may even start from scratch and regard
the \fra\ as the single input of the programme: given the \furu s,
it is possible to determine the \cdim s of all primary fields of the
theory up to a few integer constants \cite{chra,lewe,jf14}; for each
allowed set of values of these parameters one can then proceed as before.

The various types of parameters introduced here, such as the \cdim s,
the grades $\mu_m$, or the positions of \asis~and the associated
exponents, distinguish
among different \cfts~which all possess a prescribed \fra.
After imposing the polynomial equations
one is essentially left with those parameters which do not affect
the monodromy \rep\ of the relevant \deq. Two types of
such parameters can be distinguished: first,
the integer parts of the exponents (i.e., the
integer parts of \cdim s, the grades $\mu_m$, and the exponents at the
\asis), and second, the positions of the \asis. While the former are
discrete parameters, the latter are a priori continuous, and
(apart from the mild restrictions mentioned above)
the principles of rational \cft~do not seem to give
any information on them.
In particular, one can continuously deform the positions of \asis~in
such a way that the duality matrices are left invariant. It is an open
question (for more details, see \cite{jf21}) whether this means
that to any allowed value of these positions there corresponds a consistent
\cft, which would imply the existence of
continuous families of rational \cfts.
\\[11 mm] {\bf Acknowledgement.} It is a pleasure to thank Christoph
Schweigert for many helpful discussions and for a critical reading of the
manuscript.

 \end{document}